\documentclass[sigconf]{acmart}

\renewcommand\footnotetextcopyrightpermission[1]{} 

\AtBeginDocument{%
  \providecommand\BibTeX{{%
    \normalfont B\kern-0.5em{\scshape i\kern-0.25em b}\kern-0.8em\TeX}}}

\usepackage{stfloats}
\usepackage{url}

\usepackage{amsmath,amsthm,multirow,array,bm} 
\usepackage{hyperref}
\usepackage[capitalize]{cleveref}
\usepackage{algorithm}
\usepackage{algorithmic}
\usepackage[caption=false]{subfig}
\usepackage{graphicx}
\usepackage{caption}
\usepackage{adjustbox}
\usepackage{empheq}

\usepackage{cellspace}
\begin{document}

\title{Poisoning Online Learning Filters: DDoS Attacks and Countermeasures}



\author{Wesley Joon-Wie Tann \& Ee-Chien Chang } %
\affiliation{%
  \institution{Department of Computer Science, National University of Singapore}
  \country{}   }
\email{wesleyjtann@u.nus.edu,  changec@comp.nus.edu.sg}


\begin{abstract}
The recent advancements in machine learning have led to a wave of interest in adopting online learning-based approaches for long-standing attack mitigation issues. In particular, DDoS attacks remain a significant threat to network service availability even after more than two decades. These attacks have been well studied under the assumption that malicious traffic originates from a single attack profile.
Based on this premise, malicious traffic characteristics are assumed to be considerably different from legitimate traffic. Consequently, online filtering methods are designed to learn network traffic distributions adaptively and rank requests according to their attack likelihood. During an attack, requests rated as malicious are precipitously dropped by the filters.
In this paper, we conduct the first systematic study on the effects of data poisoning attacks on online DDoS filtering; introduce one such attack method, and propose practical protective countermeasures for these attacks. We investigate an adverse scenario where the attacker is ``crafty'', switching profiles during attacks and generating erratic attack traffic that is ever-shifting. This elusive attacker generates malicious requests by manipulating and shifting traffic distribution to poison the training data and corrupt the filters. To this end, we present a generative model \textsc{MimicShift}, capable of controlling traffic generation while retaining the originating traffic's intrinsic properties. 
Comprehensive experiments show that online learning filters are highly susceptible to poisoning attacks, sometimes performing much worse than a random filtering strategy in this attack scenario. At the same time, our proposed protective countermeasure diminishes the attack impact. 
\end{abstract}

\ccsdesc[500]{Networks~Denial-of-service attacks}
\ccsdesc[500]{Security and privacy~Network security}

\keywords{Poisoning attacks and countermeasures, Online learning, DDoS} 


\maketitle

\section{Introduction}

The accelerating adoption of machine learning in the enterprise makes poisoning attacks~\cite{10.5555/3157096.3157308,yang2017fake,jagielski2018manipulating} on learning systems increasingly common. These attacks generally compromise the integrity of a learning system by injecting unsanitized malicious samples into the training dataset, leading to terrible system performance. As a result, it poses a serious problem for online cyberattack mitigation systems.
One such cyberattack, the Distributed Denial-of-Service (DDoS), denies legitimate users access to resources on a target network by sending an overwhelmingly large amount of traffic to
exhaust the victim network's resources. As a result, legitimate users receive severely degraded to no service. As recent as Q1 2020,  Amazon Web Services (AWS) reported the largest publicly disclosed attack on record at 2.3 Tbps~\footnote{\url{https://aws-shield-tlr.s3.amazonaws.com/2020-Q1_AWS_Shield_TLR.pdf}}. The previous record was held by the GitHub Memcached-based DDoS attack, which peaked at 1.35 Tbps in February 2018.

As more and more present-day adaptive DDoS filtering methods adopt an online machine learning approach, we anticipate the growing urgency to study the effects of poisoning attacks on DDoS filters and find practical defenses against them. 
Online approaches typically perform incremental updates as more attack requests are observed to mitigate the malicious requests rapidly during an attack. However, such online learning methods assume that the attackers are static and that their behaviors stay the same during an attack. While these approaches are applicable to a static attacker, what if ``crafty'' attackers can dynamically change their attack behavior? We focus on studying such shifting attacks on online learning-based DDoS filtering.
Existing online learning-based filtering mechanisms can be broadly categorized into two groups. In one group, the approaches~\cite{7987186,lima2019smart,DANESHGADEHCAKMAKCI2020102756} either employ simple decision tree methods or basic machine learning algorithms. The other group~\cite{wjt_asiaccs21,Doriguzzi_Corin_2020} takes a deep learning approach, where the latest work~\cite{wjt_asiaccs21} proposed two online filtering methods designed to be computationally efficient to respond in a real-time environment where time constraint is a crucial consideration. 
Conversely, there is minimal literature that analyzes the effectiveness of fluctuating DDoS attacks on filtering methods. One study~\cite{1498362} found that adaptive statistical filters cannot effectively mitigate malicious traffic from attackers that dynamically change their behavior, performing much worse against a dynamic attacker than if the attacker was static. Most other works~\cite{ibitoye2019analyzing,8466271,10.1145/3359992.3366642,8995318,8995318,9288358,9344707} perform adversarial attacks on targeted DDoS detection systems. However, there is no study on the effects of erratic attack traffic on online learning DDoS filters.

In this paper, we study the effects of poisoning attacks on online learning-based filtering, where the attacker dynamically changes its behavior and generates fluctuating requests that vary from one time interval to the next. Our investigation of the elusive DDoS attack scenario is much more elaborate than those considered in existing online filtering approaches~\cite{wjt_asiaccs21,7987186,lima2019smart,DANESHGADEHCAKMAKCI2020102756}. In particular, we study a sophisticated setting where the attacker is ``crafty'', shifting its attack behavior (e.g., distribution of requests and features) periodically to corrupt the adaptive filters. 
The attacker models the attack traffic 
and generates malicious traffic $\mathcal{A}_t$ with different distributions in short timeframes $t$'s while mimicking properties of normal traffic. 

        \begin{figure}[tbp]
        \centering
        \includegraphics[width=1\linewidth]{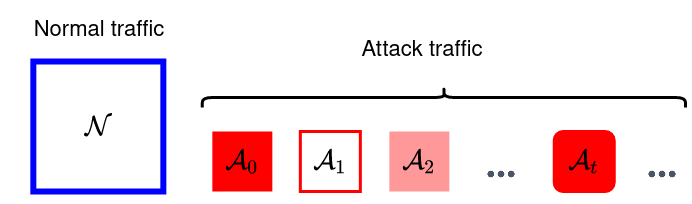}
        \caption{\textbf{Illustration of fluctuating DDoS traffic}: In this case, a ``crafty'' attacker is not only able 
        to generate attack traffic that mimics characteristics of normal traffic; the attacker constantly shifts the attack traffic $\mathcal{A}_t$ distribution from one interval to the next, in order to corrupt and elude online learning filtering methods.
        }
        \label{fig:casestudy}
    \end{figure}    


Under this shifting attack setting (see Figure~\ref{fig:casestudy}), the online learning filters utilize traffic data from two operational periods, (1) normal-day traffic $\mathcal{N}$, and (2) the attack-day traffic $\mathcal{A}$, to learn and update its filtering policy. Normal traffic behavior can be learned by performing intensive processing on $\mathcal{N}$. However, during an attack, the incoming traffic $\mathcal{A}$ is divided into short intervals of $\ell$ minutes (e.g., $\ell=1$). The attack traffic ($\mathcal{A}_0, \mathcal{A}_1, \mathcal{A}_2, \ldots$), where $\mathcal{A}_t$ contains the traffic during the $t$-th interval, will be used for the online training and update of the filters to learn attack traffic behavior. The filters then use these last updated models to create and identify a blacklist of attack identities (e.g., IP addresses) and filter them accordingly.

To this end, we formulate a controllable generative method that captures the sophisticated attack setting described above. We present \textsc{MimicShift}, a conditional generative method for controlling network traffic generation to generate shifting and unpredictable traffic. Our method is inspired by a controllable approach~\cite{tann2021rm} that is based on a conditional generative adversarial network, and it addresses the challenge of generating network requests of varying distributions. It is achieved by using request features such as \texttt{Request Len} and \texttt{TCP Window Size} that capture the intrinsic request properties. These features form the condition that we control to guide the traffic generative process. The method introduces control by leveraging the conditions, managed with a transparent Markov model, as a control vector. It allows us to influence the generative process. By specifying the density distribution parameters of the Markov model, the method can generate requests with different proprieties.

We show that the attacker can effectively inundate online learning filtering systems through the clever shifting of attack distribution periodically through such poisoning attacks.
We maintain that it is a compelling result as it suggests that existing proposed online filtering mechanisms are inadequate in mitigating poisoning attacks against online learning systems. We design a deep-learning-based generative approach, \textsc{MimicShift}, and demonstrate the effectiveness of poisoning attacks. The main advantage of \textsc{MimicShift} is the controllability of the generative process. 
It first learns the behavior of given traffic (either malicious traffic of a single attack profile or benign traffic of the victim that has been compromised) and generates synthetic traffic that is subject to manipulation, producing requests of different specified distributions. 
The control is performed in an easy-to-understand Markov process to derive traffic distribution based on the Markov model parameters. We believe that this Markov process provides an intuitive approach to the control of shifting traffic distributions.

We evaluate the proposed controllable poisoning attack on two state-of-the-art online deep-learning DDoS filtering methods, N-over-D and Iterative Classifier~\cite{wjt_asiaccs21}. 
Our evaluation is performed on the same datasets presented in the existing online filtering work~\cite{wjt_asiaccs21} to set a fair comparison benchmark.
The two publicly available network datasets are (1) CICIDS2017~\cite{icissp18} and (2) CAIDA UCSD ``DDoS Attack 2007''~\cite{caida07}. We use three separate DDoS attacks from these datasets, containing both generated traffic and anonymized traffic traces from a real-world attack, respectively. 

In the experiments, we first train a filtering model $N$ based on anomaly detection techniques. It uses only the normal traffic for learning to determine if requests during an attack are legitimate or malicious. This model sets a baseline for us to compare \textsc{MimicShift}.
Empirical results show that \textsc{MimicShift} can generate attack traffic that eludes online filtering systems, attaining close to 100\% false-negative rates at times (see Table~\ref{tab:benchmarkperf}, Figures~\ref{fig:fneg_graphs} and \ref{fig:fpos_graphs} for experiment details). It effectively negates the filtering capabilities, making the performance worse than a 50-50 random guess. 
Next, we introduce a simple yet effective protective countermeasure that greatly minimizes the attack impact, achieving a significant boost in filtering performance (see Table~\ref{tab:iterperf_enhanced}). It serves as a starting point for the future research of robust countermeasures.

Furthermore, our experiments also show that the different acceptance or rejection thresholds significantly affect the filtering performance. 
A slight change in the acceptance threshold could considerably affect the false-negative rate, allowing markedly more attack requests to be accepted. The result powerfully demonstrates the robustness of \textsc{MimicShift}. 
Interestingly, even when we only use a random combination of three Markov model parameter settings for the different attack traffic distributions, it severely degrades the online filtering performance. 
While this may appear counter-intuitive, we believe that two reasons can explain the results. (1) The primary objective of the \textsc{MimicShift} generative method is to shift its behavior unpredictably to maintain an obscure attack profile, and (2) the online filtering methods are designed to adapt quickly to new data and place higher importance on the latest observed attack traffic. Hence, online learning filters could be easily influenced and deceived by the shifting attack traffic.

\vspace{1mm} 
\noindent\textbf{Contribution.} 
\begin{enumerate}
\item 
We formulate the problem of corrupting and eluding online DDoS filtering by poisoning attacks. The attacker generates attacks that mimic the observed traffic and shift behaviors during the attack process. 

\item We design a mimic model, which can replicate features of any given requests, producing features of specified distributions that follow a simple and easy-to-understand Markov model. 
Leveraging on this mimic function, we propose a controllable generative attack method, which undermines the capabilities of online filtering systems, making them perform worse than a 50-50 random guess filtering strategy. 

\item We propose a simple yet practical countermeasure that dramatically diminishes the success of poisoning attacks on online learning filters.

\item We conduct experimental evaluations of the poisoning attacks on two state-of-the-art online deep-learning-based DDoS filtering approaches. The empirical results demonstrate both the efficacy and robustness of the attacks generated by \textsc{MimicShift}. 

\end{enumerate}




\section{Background and Challenges}
\label{sec:background}
This section gives an overview of existing machine learning-based DDoS filtering methods and detection mechanisms; we highlight the ease of detecting DDoS attacks in such systems. Next, we cover attacks on DDoS defenses and highlight poisoning attacks.

\subsection{Online DDoS Filtering}
Several unsupervised learning mitigation strategies~\cite{lima2019smart,7987186,DANESHGADEHCAKMAKCI2020102756} have been proposed to counter such attacks in an online manner. 
These learning-based methods, employing rudimentary machine learning algorithms, e.g., random forest, logistic regression, decision tree, perform clustering and simple distance measure using a few elementary features. While they can capture some basic network traffic properties, they are unable to learn deeper representations of legitimate and malicious requests to better distinguish between the two types of traffic. Moreover, the optimization goal of these methods is not clear.

In one of the most recent online deep learning filtering methods presented in Tann et al.~\cite{wjt_asiaccs21}, the two proposed approaches perform filtering by utilizing both the normal-day and the attack-day mixture traffic. 
One approach estimates the likelihood that each request is an attack, given that it is from the attack-day mixture traffic. 
The other approach formulates a loss function specific to a two-class classifier with the objective to classify requests from normal traffic as normal and an estimated proportion of requests during the attack as malicious.
The two approaches demonstrate to be suitable for attacks with a short time frame.

\subsection{Attacking DDoS Defenses}
While several DDoS filtering methods exist, there is a shortage of literature on deep generative models targeting DDoS detection systems. In recent years, some methods~\cite{ibitoye2019analyzing,8466271} have been proposed to generate adversarial samples to evade DDoS detection. On the one hand, some of these methods were employed in a white-box setting. The attackers are assumed to know the details of the target deep-learning intrusion detection system, which is not practical in the real world. 
On the other hand, another group of methods~\cite{10.1145/3359992.3366642,8995318,8995318,9288358,9344707} is presented as black-box attacks, where the attacker has very limited knowledge of the target model, such as only the model's output. Both evasion methods perform perturbations on samples until an adversarial sample successfully evades the target DDoS detector.

While these methods perform evasive attacks on DDoS intrusion systems by manipulating adversarial samples in one way or another, there is no study on the effectiveness of poisoning attacks on online deep-learning DDoS filters.
The most relevant work is a study~\cite{1498362} on adaptive statistical filters. It was determined that adaptive statistical filters perform much worse against attackers that dynamically change their behavior. However, there are no such studies for online deep learning filters.

\subsection{Data Poisoning}
Data poisoning is a type of attack, where an attacker can inject malicious data that target machine learning models during the training phase. 
Most existing poisoning attacks~\cite{10.5555/3042573.3042761,jagielski2018manipulating,rubinstein2009antidote,xiao2015feature,yang2017fake} compromise the training dataset.
The attacks aim to compromise the integrity of a machine learning system by having a model learn on unsanitized data, which manipulates a model to learn a high error rate on test samples during operation, leading to abysmal performance and the nullification of the model's effectiveness. It was first studied by Biggio et al.~\cite{10.5555/3042573.3042761} against support vector machines. Beyond linear classifier models, others have studied poisoning in areas such as collaborative filtering~\cite{10.5555/3157096.3157308}, generative method against neural networks~\cite{yang2017generative}, and topic modeling~\cite{pmlr-v38-mei15}. However, there is no study on data poisoning attacks targeting online learning filtering methods.

\section{Poisoning DD\lowercase{o}S Filters} 
\label{sec:formulate}   
In this section, we formulate the problem of poisoning DDoS filters and discuss the challenges of avoiding online learning detection models that constantly update themselves. We provide an example to describe the problem and highlight its relevant concepts.

\subsection{Poisoning Attack}
We model the interactions between a ``clever'' attacker and an online-learning DDoS filter as a game. In this game, there are mainly three notions relating to every time interval $t$: (1) the estimated attack traffic distribution $\widehat{A}_t$ of the filter at each interval, (2) the generated attack traffic distribution $\widetilde{A}_t$ that the attacker constantly shifts from one interval to the next, and (3) the filter model $F_{t}$, which makes the decision on requests that are to be dropped during an attack. The goal of an attacker is to shift the distribution of attack traffic dynamically, deceiving the filter into approximating an inaccurate attack distribution, such that it is unable to distinguish between normal and attack requests accurately; effectively carrying out a poisoning attack.

\vspace{3mm} 
\noindent\textbf{Motivating Example.}
Let us consider a scenario where an elusive attacker aims to disrupt the network service availability of a target victim server by overwhelming the target with a flood of unpredictable attack traffic. In this scenario, the attacker decided to attack the victim service provider in the application layer, as such attacks are especially effective at consuming server and network resources. 
A request in the application-layer DDoS could be a sequence $\bm{x} = (x_1, x_2, \ldots, x_{L})$ of sub-requests $x$'s. For example, a visit to a website consists of an \texttt{HTTP-GET} request to the landing page, followed by other requests. Each user who makes a request $\bm{x}$, a sequence of sub-requests, corresponds to an IP address. 

Most network service providers would have protective DDoS filtering mechanisms 
monitoring the availability of their networks. Such a DDoS filtering mechanism is often an online learning-based filter that makes the decision to drop abnormal requests based on some extracted salient features, with an adaptive model trained on observed network traffic. Therefore, it is reasonable to assume that the attacker would not have any knowledge about the protective filtering system that the network service provider has in place (e.g., the features, time interval, and data used by the filter). Nevertheless, the attacker can always probe the filter by making malicious requests and studying the provider's response (accept/reject) via the status code.
For example, receiving the code \texttt{HTTP 403} means that the requested resource is forbidden by the provider. However, we assume that the attacker can only probe the provider's filter a limited number of times before its IP address is blacklisted. 

To avoid detection, the attacker has a generator that generates controllable network requests that model regular traffic closely. 
For instance, the generator is able to dynamically generate malicious traffic requests by allowing the attacker to control the distribution shift of generated traffic. Due to the attacker's limited knowledge of the underlying online DDoS protection mechanism, only by constantly shifting its attack traffic distribution, can the attacker confuse the online learning filter, and it allows the attacker to be always one step ahead of the filter's estimation of the attack profile. Given the strict constraints, we aim to investigate the effectiveness of such attacks.

\subsection{Online Learning-Based Filtering}
For the online filter, we use the deep learning filters~\cite{wjt_asiaccs21} that demonstrated effectiveness against DDoS attacks in an online setting. The two filters proposed are the \textit{N-over-D} nets and \textit{Iterative Classifier}.
Their goal is to filter out an estimated proportion of malicious requests during an attack. The considered filter is one that learns on both the traffic from two operational periods for model training. As an abnormally large number of requests characterizes attacks, it can be readily detected in practice. The first period is the normal-day traffic $\mathcal{N}$, where the requests are logged during the normal period. The second period is the attack-day traffic $\mathcal{A}$ that contains a mixture of unlabeled legitimate and malicious traffic. Based on the volume of attack traffic, it is assumed that the filter is able to roughly estimate the proportion of attack requests 
in $\mathcal{A}$. All requests in $\mathcal{N}$ are considered to be normal, whereas traffic in $\mathcal{A}$ is unlabeled.

In an online setting, the filtering process can be broadly categorized into three stages. First, during a normal period, intensive processing and training can be performed on normal-day traffic $\mathcal{N}$ to learn an accurate intermediate model $\widehat{N}_0$ of normal requests. Next, when an attack begins, the filter model starts its online learning of incoming mixture traffic $\mathcal{A}$. The mixture traffic can be divided into intervals of $\ell$ minutes (e.g., $\ell=1$), and ordered accordingly ($\mathcal{A}_0, \mathcal{A}_1, \mathcal{A}_2, \ldots$), where $\mathcal{A}_t$ contains the mixture within the $t$-th interval. At each interval $t$, the model, trained on $\mathcal{A}_{t-1}$ and some intermediate representation $\widehat{N}_{t-1}$, giving the updated $\widehat{N}_{t}$ and a model $F_{t}$. 
Finally, the trained model $F_{t}$ is used to estimate the attack traffic distribution $\widehat{A}_t$ and perform the filtering of incoming requests.

\subsection{Elusive Attacker}
The attacker takes as input a request sample $\bm{x}$ from given traffic (either malicious traffic of a single attack profile or benign traffic of the victim that has been compromised) and generates another request $\widetilde{\bm{x}} = (\widetilde{x}_1, \widetilde{x}_2, \ldots, \widetilde{x}_{L})$ that mimics the traffic. 
In addition, the attacker shifts the distribution of the generated requests in order to fool the filter. It can be achieved by leveraging some data features as conditions to control the traffic generation process. 
Each request $\bm{x}$ is associated with some selected feature that is grouped into classes $k_i \in K$ (e.g., the various request payload sizes), and it corresponds to the request in the given traffic.
Through these selected features, the attacker is able to control the distribution during the generative process. We denote this time-varying distribution as $\widetilde{A}_t$. The attacker can take advantage of the useful properties of the network features to manipulate the generated samples.

For example, the attacker can use the \texttt{TCP Len} feature as one of the conditions to vary the distribution of generated requests. In one interval, the attacker can generate attack distribution $\widetilde{A}_t$ with requests of longer packet lengths. In contrast, the next interval could be generated with short packet lengths. Nevertheless, such traffic features are not fixed, and it is always up to the attacker to select from the set of all network traffic features. 
As the goal of the attacker is to generate unpredictable requests and attempt to get the online learning filter to accept as many of them as possible, it is sufficient for the attacker to mimic the traffic and find a distribution $\widetilde{A}_t$ that the filter's estimated distribution $\widehat{A}_t$ is unable to accurately capture in time.

\section{Controllable Generative Method} 
\label{sec:overview}   
In this section, we present the controllable generative method \textsc{MimicShift} that generates unpredictable malicious traffic with varying distributions for poisoning online learning DDoS detection filters. It leverages both conditional modeling and generative adversarial networks (GANs) to generate sequence data. 

    \begin{figure}[htbp]
        \centering
        \includegraphics[width=1.0\linewidth]{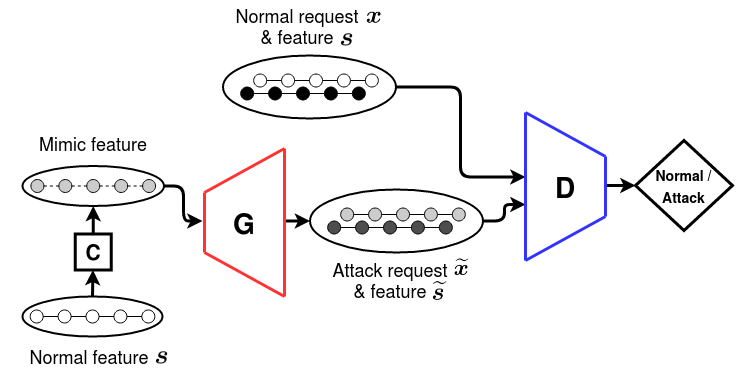}
        \caption{\textsc{MimicShift}: The proposed generative method for controlling and generating network traffic with shifting distributions.}
        \label{fig:mimicshift}
    \end{figure} 
    

    
We design \textsc{MimicShift} as a controllable network request generation method (see Figure~\ref{fig:mimicshift}). 
It is inspired by the conditional GAN framework~\cite{DBLP:journals/corr/MirzaO14}.
\textsc{MimicShift} leverages selected traffic features (e.g., \texttt{Request Len} or \texttt{TCP Window Size}) as a control vector to influence the generative process. During the training phase, it first takes given requests $\bm{x}$ as input and selects some corresponding features $\bm{s}$ as the condition for controlling the request generation. Next, the method trains an imitative mimic model $C$, which inputs the feature sequences $\bm{s}$ of the corresponding requests $\bm{x}$ in the given traffic. It then generates representative feature sequences $\widetilde{\bm{s}}$ that mimic the real features. After learning to mimic the features, a generative model $G$ then takes these representative feature sequences $\widetilde{\bm{s}}$ as conditions to produce generated attack requests $\widetilde{\bm{x}}$. The objective of $G$ is to learn the distribution of given traffic requests and generate synthetic requests that are as alike as possible to the requests. Simultaneously, a discriminative model $D$ estimates the probability that a request sequence and its corresponding features are produced by $G$ or a real request; its goal is to distinguish between the actual and synthetic requests.

Following the training of models $C$, $G$, and $D$, we specify the Markov model parameters ($\bm{\pi}, \bm{\alpha}$) for controlling the mimic model $C$, which produces representative requests based on the specified parameters to produce representative features $\widetilde{\bm{s}}$ that follow the intended transition dynamics. 
We then control the request traffic generative process with these designed features $\widetilde{\bm{s}}$, influencing the generator $G$ to create synthetic manipulated requests $\widetilde{\bm{x}}$. 

In our method, we train both $G$ and $D$ conditioned on selected features  $\bm{s}$ that correspond to the requests $\bm{x}$. We designed our model to allow itself to be significantly influenced by some user-specific features such as \texttt{TCP Len} or any other network request feature.  As a result, the model can $(1)$ introduce a high dimensional control vector to generate unpredictable requests, and $(2)$ directly control the data generation process to confuse and elude DDoS filters. For example, by using payload information of the requests, we can learn the distribution of regular payloads, using this information, then explicitly generate requests that mimic the traffic closely. 

Based on the first introduced conditional GAN~\cite{DBLP:journals/corr/MirzaO14}, we present the conditional GAN training for network traffic requests, defining the loss function as:
\begin{equation} \label{eq:cganloss}
    \mathcal{L} = \log (D(\bm{x} \mid \bm{s})) + \log (1 - D(G(\bm{z} \mid \bm{s})))
\end{equation}   

where $\bm{z} \sim \ \mathcal{N}(\bm{0}, \bm{I}_d)$ is the latent noise sampled from a multivariate standard normal distribution. 
The network traffic is represented as a binary adjacency matrix $\bm{\bar{A}} \in \{0,1\}^{N \times N}$ of $N$ different request types. We then sample sets of random walks of length $T$ from adjacency matric $\bm{\bar{A}}$ to use as training data $\bm{x}$ for our model. 
Following a biased second-order random walk sampling strategy, the sampled random walks are network requests where each request could be made up of a sequence $\bm{x} = (x_1, x_2, \ldots, x_{L})$ of sub-requests $x$'s. 
One desirable property of such a random walk is that the walks only include connected sub-requests, which efficiently exploits the sequential nature of requests in real-world network traffic.

\section{Implementation}
\label{sec:details}
In this section, we describe each component of the \textsc{MimicShift} method and formally present the controllable generative process.

\subsection{Mimic} 
The mimic model $C$ is  
designed to learn traffic behavior by modeling features of sub-requests in the sampled requests. It is a sequence-to-sequence model.
Given a sequence of traffic request features, the model then predicts the input sub-request features one at a time as the features are observed. We design $C$ using a long short-term memory (LSTM)~\cite{10.1162/neco.1997.9.8.1735} neural network. 
The purpose of $C$ is to model network traffic, retain a memory of its behavior, and provide a vector of control for the ensuing generative process.
When given sampled sequences of features $(\bm{s}_1, \hdots, \bm{s}_{L})$ as inputs, the mimic $C$ learns to generate synthetic features $(\widetilde{\bm{s}}_1, \hdots, \widetilde{\bm{s}}_{L})$ that mimic the input features. For example, a sample input \texttt{Request Len} sequence $(60, 1514, 52, 60, \hdots, 66)$ given to $C$ could result in such an output of $(60, 60, 1514, 60, \hdots, 66)$, which closely mimics the input sample. 

\subsection{Generator}
The generator $G$ generates network requests $(\widetilde{\bm{x}}_1, \hdots, \widetilde{\bm{x}}_L) \sim G$ conditional on the selected features. 
It is a parameterized LSTM model $f_\theta$. 
At each step $l$, the LSTM $f_\theta$ takes as input the previous request $\widetilde{\bm{x}}_{l-1}$, the last memory state $\bm{m}_{l-1}$ of the LSTM model, and the current feature $\widetilde{\bm{s}}_l$.
The model then generates two values $(\bm{p}_l, \bm{m}_l)$, where $\bm{p}_l$ denotes the probability distribution over the current request and $\bm{m}_l$ the current memory state. Next, the generated request $\widetilde{\bm{x}}_l$ is sampled from a categorical distribution $\widetilde{\bm{x}}_l \sim Cat(\sigma(\bm{p}_l))$ using a one-hot vector representation, where $\sigma(\cdot)$ is the softmax function.
%
A latent noise from a multivariate standard normal distribution $\bm{\bm{z}} \sim \ \mathcal{N}(\bm{0}, \bm{I}_d)$ is drawn and passed through a hyperbolic tangent function $g_{\theta'}(\bm{\bm{z}})$ to compute the initial memory state $\bm{m}_0$. 
Generator $G$ takes this noise $\bm{\bm{z}}$ and sampled features $\bm{s}$ as inputs, and generates attack requests $(\widetilde{\bm{x}}_1, \hdots, \widetilde{\bm{x}}_L)$. 

\subsection{Discriminator}
The discriminator $D$ aims to discriminate between requests sampled from the given traffic and synthetic requests generated by the generator $G$. It is a binary classification LSTM model that takes two inputs: the current request $\bm{x}_l$ and its associated feature condition $\bm{s}_l$, both represented as one-hot vectors. 
$D$ then outputs a score between $0$ and $1$, indicating the probability of an actual request.


\subsection{Controlling Traffic Generation} 
\textsc{MimicShift} takes traffic features as conditions to input into the generator and influences the generated output requests, creating requests with various feature densities. Our method constantly shifts the distribution of these conditions to produce unpredictable network traffic. The method aims to provide controllability to the attack and generate requests that mislead the filter, allowing the attack requests to elude detection. 
%
We first build a Markov chain model according to some user-specified transition distribution and construct specific features sequences for requests of some desired distribution. These sequences are then injected into the mimic $C$ to generate features $(\widetilde{\bm{s}}_1, \hdots, \widetilde{\bm{s}}_{L})$ that mimic the sampled features of given requests.
Next, we train the \textsc{MimicShift} model 
on given requests $\bm{x}$. Given the trained model, we then construct features
that follow some specified distribution and use the generator $G$ to produce desired requests $\widetilde{\bm{x}}$. Through this process, the attacker can control the feature conditions and generate unpredictable requests with varying distributions.

For example, by controlling the parameters of the mimic feature, our method has the ability to generate attack traffic that has shifting distributions. The features provide a vector of control to how requests are generated. We influence the generative process by constructing mimic features of desired distribution using the mimic model $C$. 
First, we construct select features by specifying the parameters of a transparent Markov model: (1) initial probability distribution over $K$ classes $\bm{\pi} = (\pi_1, \pi_2, \dots , \pi_{K})$, where $\pi_i$ is the probability that the Markov chain will start from class $i$, and (2) transition probability matrix $\bm{\alpha} = (\alpha_{11}\alpha_{12}\dots \alpha_{k1}\dots \alpha_{kk})$, where each $\alpha_{ij}$ represents the probability of moving from class $i$ to class $j$.
Next, we input these constructed sequences into mimic $C$, and it returns model-generated features that replicate the original features. Finally, by injecting these constructed sequences into our trained generator $G$, we generate attack requests that follow the specific distribution.

\section{Evaluation}
\label{sec:eval}
In this section, we present the evaluation of online deep learning filters performed on various DDoS attacks with fluctuating requests in two scenarios, where (1) the attacker has access to the regular traffic of the victim that has been compromised (with normal traffic $\mathcal{N}$), and (2) the attacker uses traffic from a single attack profile to approximate regular traffic (without normal traffic $\mathcal{N}$).
We demonstrate the ineffectiveness of such filters against our generative attack method \textsc{MimicShift}, and compare their performance against known DDoS attacks.
Finally, we analyze the effects of such poisoning attacks and introduce an effective countermeasure.

\subsection{Datasets}
\label{subsec:datasets}
Our evaluation is performed on the same attacks presented in the existing online DDoS filtering paper~\cite{wjt_asiaccs21} to set a fair comparison benchmark. The datasets are from two publicly available network datasets: CICIDS2017~\footnote{www.unb.ca/cic/datasets/ids-2017.html}~\cite{icissp18} and CAIDA UCSD ``DDoS Attack 2007''~\footnote{www.caida.org/data/passive/ddos-20070804\textunderscore dataset.xml}~\cite{caida07},
containing generated network traffic resembling real-world attacks and anonymized traffic traces from a DDoS attack on August 04, 2007, respectively. From these two datasets, three distinct DDoS attacks are extracted for evaluation.

\vspace{3mm} 
\noindent\textbf{Synthetic Benchmark Data.}
In the first dataset for evaluation, we use the synthetic \textit{CICIDS2017}~\cite{icissp18} dataset provided by the Canadian Institute for Cybersecurity of the University of New Brunswick (UNB), Canada. The human traffic was generated by modeling the abstract behavior of user interactions, and the response of 25 users based on the HTTP, HTTPS, FTP, SSH, and email protocols were used to create the abstract behavior. 
The two selected attacks: (1) HULK and (2) Low Orbit Ion Canon (LOIC) last for 17 and 20 mins, respectively. 

\vspace{3mm} 
\noindent\textbf{Real-world Data.}
In the second dataset, we use a real-world attack, \textit{CAIDA07} UCSD ``DDoS Attack 2007''~\cite{caida07}. 
This attack occurred on August 04, 2007, and affected various locations around the world. 
It contains anonymized traffic traces from a DDoS attack that lasted around an hour, including network traffic such as Web, FTP, and Ping. 
These traces mostly only include attack traffic to the victim and responses from the victim, and non-attack traffic has been removed as much as possible. The network load surges dramatically from about 200 kbits/s to about 80 Mbits/s within a few minutes. 

\subsection{Mimic Attacks}
Next, we generate fluctuating DDoS attack traffic based on both the \textit{CICIDS2017} and \textit{CAIDA07} datasets to perform a detailed evaluation of the online learning filtering methods on these attacks. First, we select the feature to mimic and use as a control vector to influence the generative process. After selecting the appropriate feature, we specify the distributions for shifting the attack requests, and generate fluctuating traffic to attack the online filtering methods. 

\vspace{3mm} 
\noindent\textbf{Feature Analysis.}
In order to select a feature for the appropriate condition as a control vector, we perform an analysis of the features in the various datasets. We found that the distributions of the features of the traffic are highly skewed. This skewness is because a large majority of the traffic is concentrated at a few top features, such as the most common IP flags, TCP lengths, and window sizes.

    \begin{table}[htbp]
  \centering
  \caption{Feature analysis of \texttt{Request Len} in the traffic of all three datasets. The distribution of the requests are mostly concentrated in the top few classes of the feature.}
\resizebox{0.7\linewidth}{!}{%
\begin{tabular}{
    >{\centering\arraybackslash}m{1.5cm}
    |>{\centering\arraybackslash}m{1.5cm}
    >{\centering\arraybackslash}m{.5cm}
    >{\centering\arraybackslash}m{3cm}
    }
    \hline
\textbf{Dataset} & \textbf{\# \texttt{Request Len}} &\textbf{Top 3 (\%)} & Distribution\\ \hline
\textit{HULK}       &3753 &79.7 &\includegraphics[width=.175\textwidth]{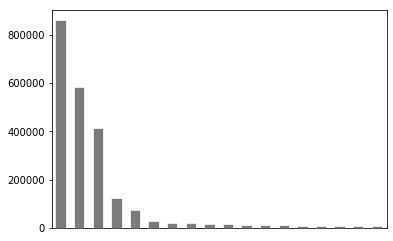}  \\ \hline
\textit{LOIC}       &3490 &60.4 &\includegraphics[width=.175\textwidth]{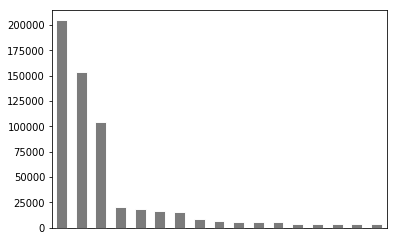}  \\ \hline
\textit{CAIDA07}    &1059 &91.2 &\includegraphics[width=.175\textwidth]{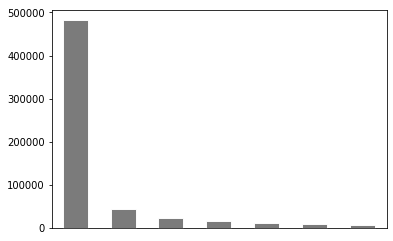}  \\ \hline
\end{tabular}
}
\label{tab:featureanalysis}
\end{table}


For example, using the \texttt{Request Len} feature as an illustration, it can be seen in Table~\ref{tab:featureanalysis} that the traffic is concentrated in the top few lengths of the feature. 
In the \textit{CAIDA07} dataset, the top three request lengths (60, 52, 1500) make up 91\% of the traffic, with the top length (60) accounting for 81\% of the traffic alone. As for the \textit{LOIC} dataset, the top three request lengths (60, 66, 1514) make up 60\% of the traffic; the top three features (60, 1514, 2974) of the \textit{HULK} dataset make up 80\% of the total traffic.

As the distribution of the feature \texttt{Request Len} is concentrated in the top few classes, we group all of them into three classes for our experiments. The first two classes are the top two features, and the third class consists of all the other features. We design it such that the distribution of the three classes is more uniform and increases the variability of feature distribution.

\vspace{3mm} 
\noindent\textbf{Distribution Parameters.}
We specify the Markov model parameters, initial probability distribution $\bm{\pi} = (\pi_1, \pi_2, \pi_3)$ and transition probability matrix $\bm{\alpha} = (\alpha_{11}\alpha_{12}\dots \alpha_{31}\dots \alpha_{33})$, used in our experiments for controlling the shift of the generated mimic attacks. We use various parameters for all the datasets (see Table~\ref{tab:markovparam}). 

    \begin{table}[htbp]
  \centering
  \caption{Generation parameters for the \textit{HULK}, \textit{LOIC}, and \textit{CAIDA07} attacks. 
  }
\resizebox{.6\linewidth}{!}{%
\begin{tabular}{|Sc|Sc|Sc|}
\hline
\textbf{Attack} & $\bm{\pi}$ &$\bm{A}$ \\ \hline
$\mathcal{A}_0$       &$\begin{bmatrix}0.9 & 0.05 & 0.05\\\end{bmatrix}$ &$\begin{bmatrix}
0.98 &0.01 &0.01 \\
0.1  &0.6  &0.3 \\
0.0  &0.1  &0.9
\end{bmatrix}$   \\ \hline
$\mathcal{A}_1$       &$\begin{bmatrix}0.05 &0.05 &0.9\\\end{bmatrix}$ &$\begin{bmatrix}
0.9 &0.1 &0.0 \\
0.1 &0.6 &0.3 \\
0.03 &0.02 &0.95
\end{bmatrix}$   \\ \hline
$\mathcal{A}_2$    &$\begin{bmatrix}0.05 & 0.9 & 0.05\\\end{bmatrix}$ &$\begin{bmatrix}
0.9 &0.1 &0.0 \\ 
0.1 &0.7 &0.2 \\
0.0 &0.1 &0.9
\end{bmatrix}$    \\ \hline
\end{tabular}
}
\label{tab:markovparam}
\end{table}


The attack is divided into three traffic ($\mathcal{A}_0, \mathcal{A}_1, \mathcal{A}_2$) with different distributions. At each time period during the attack, one of the attack parameter settings is picked at random for $\mathcal{A}_t$. Our model \textsc{MimicShift} would then generate attack traffic according to the specified distributions, and produce fluctuating traffic to attack the online filtering methods. 
\begin{table*}[!ht]
    \centering
    \captionof{table}{[\textbf{Filtering Performance}] The performance of N-over-D and Iterative Classifier on the various shifting attack traffic in \textit{CICIDS2017} and \textit{CAIDA07} datasets.}
    \resizebox{0.9\linewidth}{!}{%
        \begin{tabular}{c||c|cccccc|cccccc}
        \hline
        \multirow{2}{*}{\textbf{Dataset}} & \multirow{2}{*}{\textbf{Model}} &\multicolumn{6}{c}{Attacker with normal traffic $\mathcal{N}$} &\multicolumn{6}{c}{Attacker without normal traffic $\mathcal{N}$} \\ \cline{3-14}
        & &\textbf{FNR} &\textbf{FPR} &\textbf{ACC} &\textbf{Prec.} &\textbf{Rec.} &\textbf{F1}            &\textbf{FNR} &\textbf{FPR} &\textbf{ACC} &\textbf{Prec.} &\textbf{Rec.} &\textbf{F1} \\ \hline 

        \multirow{6}{*}{HULK} &N  
        &0.7824 &0.2790 &0.9019  &0.9196  &0.9561  &0.9375 
        &0.7834	&0.2734	&0.9045	&0.9212	&0.9578	&0.9392\\ 
        &N-over-D ($\ell=1$)      &0.8997  &0.6295  &0.7403  &0.8186  &0.8511  &0.8345
        &0.9009	&0.6432	&0.7340	&0.8147	&0.8470	&0.8305\\
        &N-over-D ($\ell=\infty$) &0.9693  &0.6461  &0.6461  &0.7597  &0.7899  &0.7745
        &0.9697	&0.8411	&0.6427	&0.7576	&0.7877	&0.7724\\
        &Iterative Classifier ($\ell=1$)  &0.6802  &0.5684  &0.8435  &0.8780   &0.9252  &0.9010
        &1.0000	&0.6772	&0.6514	&0.8326	&0.6846	&0.7514\\ 
        &Iterative Classifier ($\ell=\infty$)   &0.7006 &0.6619 &0.7864  &0.8702  &0.8548  &0.8624
        &0.2169	&0.4329	&0.7930	&0.8786	&0.8482	&0.8631\\ 
        \hline

        \multirow{6}{*}{LOIC} &N  
        &0.7863 &0.2350 &0.9062 &0.9411 &0.9416 &0.9414
        &0.7869	&0.2434	&0.9028	&0.9390	&0.9395	&0.9393\\  
        &N-over-D ($\ell=1$)      &0.8882 &0.6532 &0.7386 &0.8364 &0.8368 &0.8366   
        &0.9269	&0.7857	&0.6855	&0.8032	&0.8036	&0.8034\\  
        &N-over-D ($\ell=\infty$) &0.9836 &0.9741 &0.6100 &0.7560 &0.7564 &0.7562
        &0.9877	&0.9796	&0.6078	&0.7546	&0.7550	&0.7548\\
        &Iterative Classifier ($\ell=1$)   &0.8823 &0.2003 &0.2085 &0.0520 &0.5538 &0.0951
        &1.0000	&0.8364	&0.8330	&0.8272	&1.0000	&0.9054\\ 
        &Iterative Classifier ($\ell=\infty$)   &0.6943 &0.4102 &0.3798 &0.3129 &0.7795 &0.4465
        &0.9265	&0.2514	&0.3402	&0.8042	&0.2312	&0.3591\\ \hline

        \multirow{6}{*}{CAIDA07} &N 
        &0.7555 &0.0013 &0.9825 &0.9987 &0.9798 &0.9891
        &0.7592	&0.0155	&0.9598	&0.9845	&0.9659	&0.9751\\  
        &N-over-D ($\ell=1$)        &0.7622 &0.8025 &0.6884 &0.8149 &0.7995 &0.8071
        &0.7849	&0.0326	&0.9324	&0.9674	&0.9491	&0.9582\\ 
        &N-over-D ($\ell=\infty$)   &0.9288 &0.7905 &0.6928 &0.8177 &0.8022 &0.8099
        &0.7882	&0.0355	&0.9278	&0.9645	&0.9463	&0.9553\\  
        &Iterative Classifier ($\ell=1$)   &0.9434 &0.1109 &0.8818 &0.8754 &0.9773 &0.9236
        &0.0317	&0.9978	&0.8005	&0.8132	&0.9806	&0.8891\\  
        &Iterative Classifier ($\ell=\infty$)   &0.9266 &1.0000 &0.9753 &0.9998 &0.9708 &0.9851
        &0.8881	&0.0654	&0.3140	&0.9255	&0.1727	&0.2911\\ \hline
        \end{tabular}
    }
    \label{tab:benchmarkperf}
\end{table*} 
\subsection{Online Filtering Methods}
We perform a detailed evaluation of the online learning filtering methods on the generated mimic attacks, and report the filtering capabilities to identify attack requests from normal traffic. The two online filtering methods are the latest state-of-the-art deep learning approaches to DDoS filtering~\cite{wjt_asiaccs21}: 
\begin{enumerate}
    \item N-over-D approach
    \item Iterative classifier approach
\end{enumerate}
that take a principled approach to estimate conditional probabilities and an online iterative two-class classifier design for a specifically formulated machine learning loss function, respectively. 

In the N-over-D approach, the technique approximates the likelihood that a network request is an attack, given that it is observed in the attack traffic. It is achieved with a two-step process. The first step applies an unsupervised learning model on the normal-day traffic to obtain a model that learns the distribution of normal traffic. Then, when a DDoS attack is detected, the second step is activated, and it trains another unsupervised learning model to rapidly learn the distribution of the attack traffic. At any time, the conditional probability can be estimated by taking the ratio of the normal-traffic distribution over the attack-traffic distribution. 
As for the Iterative Classifier approach, a particular machine learning loss function is formulated for fast online learning using one learning model. 
An online iterative two-class classifier performs joint learning on both the normal and attack operational periods, fully leverages all available information for model training. 

Both filtering approaches are evaluated using the online interval of 1 min ($\ell=1$) and a rejection rate of $80\%$. Under such settings, it means that each $\mathcal{A}_t$ is collected over a period of 1 min for learning and filtering, dropping $80\%$ of the requests. 
In addition, we also use an offline approach for comparison. In this setting, the interval is set as $\ell=\infty$. We suppose that all normal and attack data is available for training the models. The offline setting effectively shows that it is not easy for models to learn to generalize over the different distributions. By taking an offline approach that trains on all available data, we can (1) study how effective the varying distributions of attack traffic are on online filtering methods, and (2) obtain a reference performance limit for the online filtering results.

\subsection{Evaluation Results}
\label{sec:eval_results}
\begin{figure*}[!htp]
    \caption{[\textbf{False Negative Rate (Attack with $\mathcal{N}$)}] The x-axis is the acceptance ratio (0--1), and the y-axis is the corresponding false accept rate. If the acceptance ratio is at 0.2, it means that only 20\% of the requests are accepted, and the graph shows the rate of false negatives.}
    \centering
    \resizebox{0.95\linewidth}{!}{
        \begin{minipage}[b]{.33\textwidth}
        \centering
        \subfloat[][HULK (FNR)]{
            \includegraphics[width=1.0\linewidth]{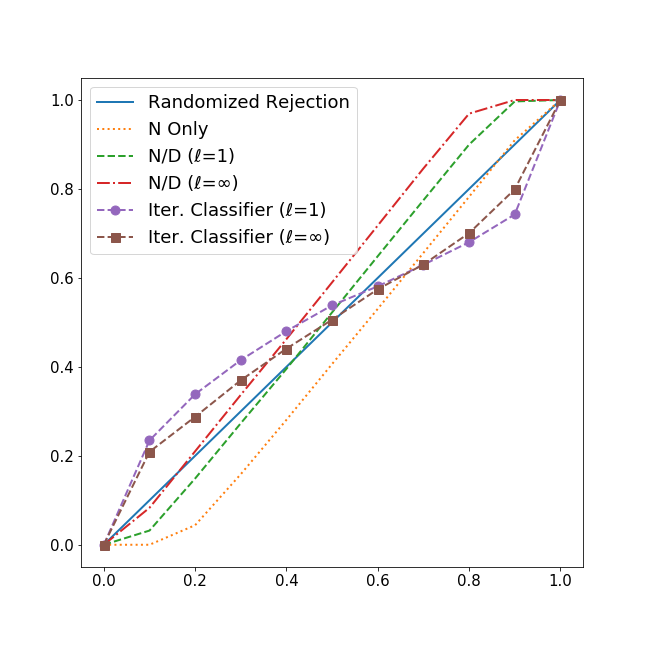}}
        \end{minipage} 
        \begin{minipage}[b]{.33\textwidth}
        \centering
        \subfloat[][LOIC (FNR)]{
            \includegraphics[width=1.0\linewidth]{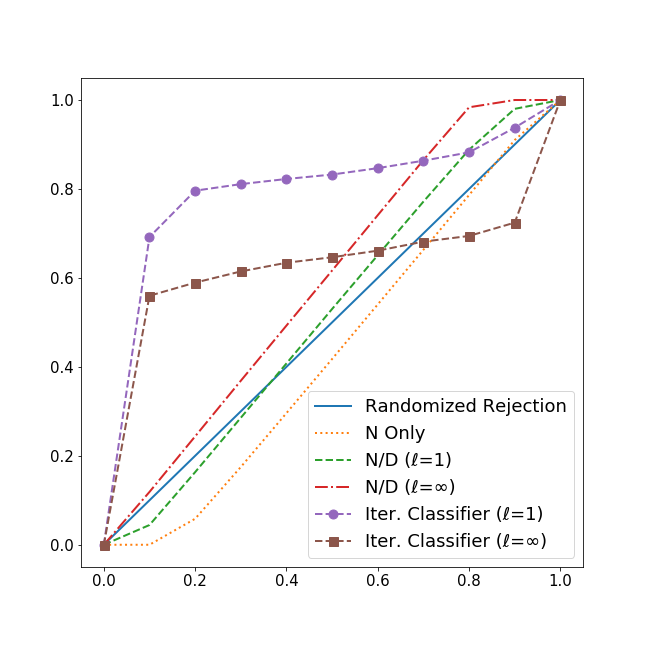}}
        \end{minipage} 
        \begin{minipage}[b]{.33\textwidth}
        \centering
            \subfloat[][CAIDA07 (FNR)]{
            \includegraphics[width=1.0\linewidth]{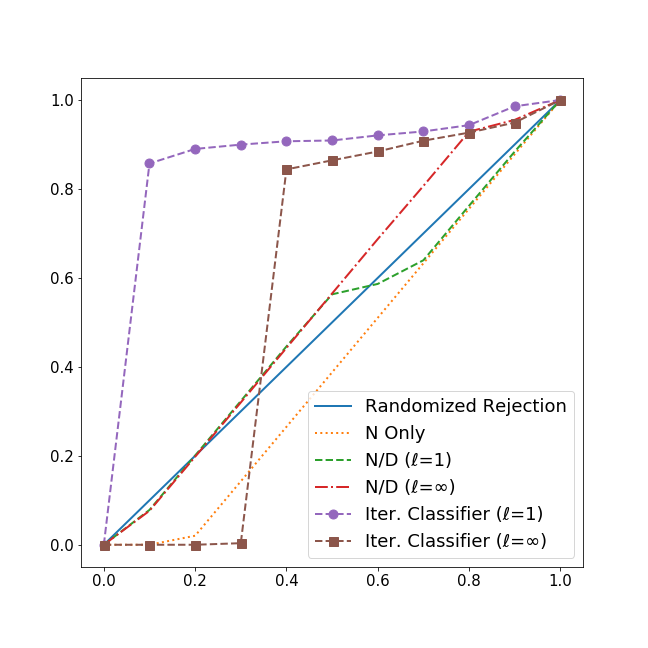}}
        \end{minipage} 
    }
    \label{fig:fneg_graphs}
\end{figure*}

\begin{figure*}[!htp]
    \centering
    \resizebox{0.95\linewidth}{!}{
        \begin{minipage}[b]{.33\textwidth}
        \centering
        \subfloat[][HULK (FPR)]{
            \includegraphics[width=1.0\linewidth]{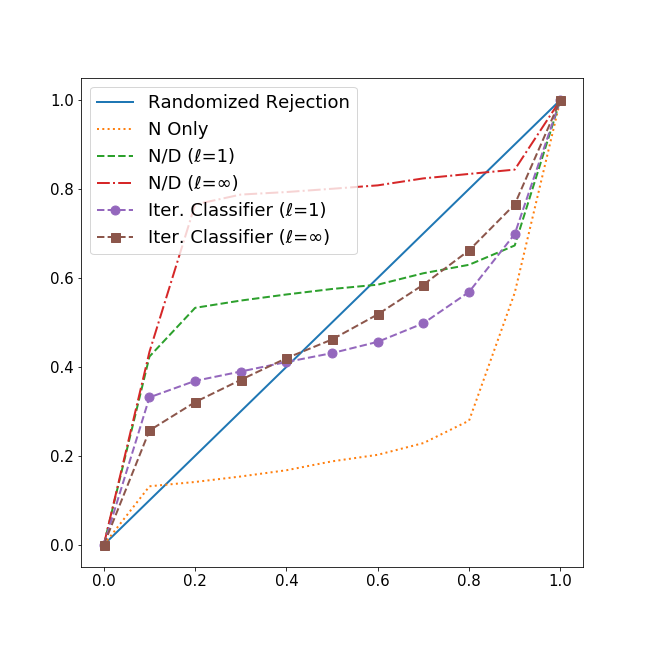}}
        \end{minipage} 
        \begin{minipage}[b]{.33\textwidth}
        \centering
        \subfloat[][LOIC (FPR)]{
            \includegraphics[width=1.0\linewidth]{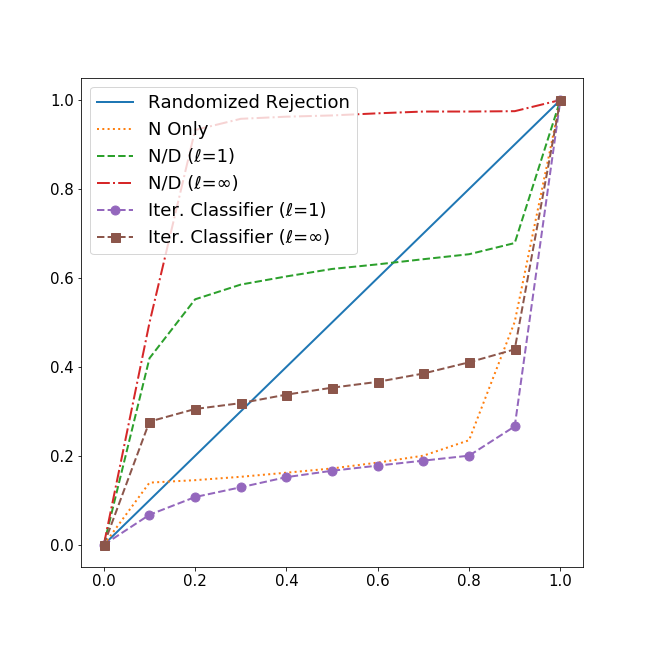}}
        \end{minipage} 
        \begin{minipage}[b]{.33\textwidth}
        \centering
            \subfloat[][CAIDA07 (FPR)]{
            \includegraphics[width=1.0\linewidth]{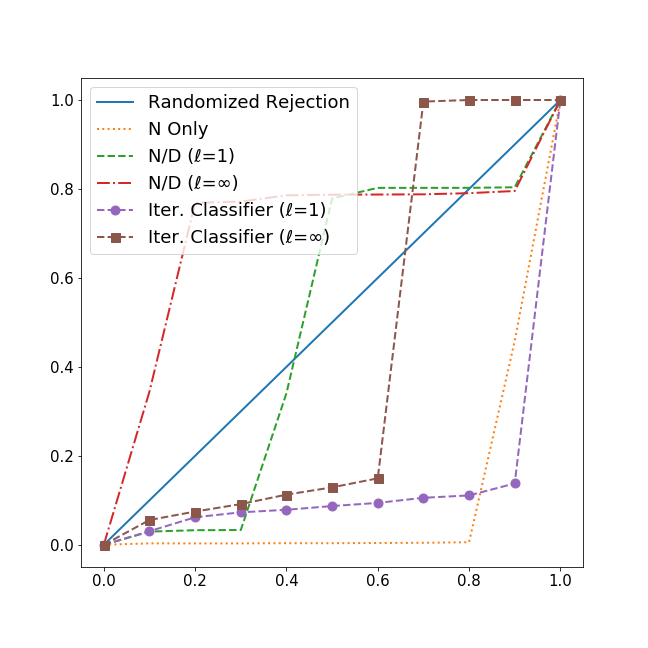}}
        \end{minipage} 
    }
    \caption{[\textbf{False Positive Rate} (Attack with $\mathcal{N}$)] The x-axis is the rejection ratio (0--1), and the y-axis is the corresponding false rejection rate. If the rejection ratio is at 0.8, it means that 80\% of the requests are rejected, and the graph shows the rate of false positives.}
    \label{fig:fpos_graphs}
\end{figure*}

We measure the performance of the two filtering approaches in both attack scenarios. As demonstrated in Table~\ref{tab:benchmarkperf}, Figure~\ref{fig:fneg_graphs}, and Figure~\ref{fig:fpos_graphs}, both N-over-D and Iterative Classifier result in high false-negative rates and high false-positive rates across the three different attacks, thereby undermining the effectiveness of the filter and reducing it to no better than a random guess. 
We present both the false-negative and false-positive graphs of the attack with normal traffic $\mathcal{N}$ as both attack scenarios have similar results (see Appendix for the \textit{attack without $\mathcal{N}$} plots).
These results show that unpredictable attack traffic essentially disables online learning filters.

The reported evaluation metrics used are False Negative Rate (FNR), False Positive Rate (FPR), Accuracy (ACC), Precision (Prec.), Recall (Rec.), and F1 Score (F1). 
The two most relevant metrics are FNR and FPR. False Negative Rate is the rate that attack requests are wrongly classified as legitimate and not dropped by the filtering methods. False Positive Rate reflects the rate that requests are from normal traffic but predicted as attack and rejected by the filters.


    %


In Table~\ref{tab:benchmarkperf}, the results demonstrate a trade-off between letting attack requests through and denying legitimate requests. The N-over-D approach achieves a lower false-negative rate (FNR) than the Iterative Classifier, but it has a much higher false-positive rate (FPR). On the one hand, it means that while the N-over-D allows a smaller proportion of the attack requests through, it rejects many of the normal requests, denying a large portion of the legitimate requests. On the other hand, the iterative classifier has a higher FNR but a lower FPR, indicating it allows more legitimate requests at the expense of more attack requests. In general, although the attacker with normal traffic $\mathcal{N}$ fares slightly better than an attacker without $\mathcal{N}$ (higher error rates), the FNR and FPR in both attack scenarios are similar; results suggest that the attacks using representative traffic can be successful.

\subsection{Result Comparison}
\label{sec:eval_compare}
For comparison, we assess the performance of state-of-the-art online DDoS filtering methods, N-over-D and Iterative Classifier approaches, and note the difference. Using the \textit{CICIDS2017} and \textit{CAIDA07} datasets, we study how our generated fluctuating attack traffic affects the filtering performance. More importantly, we demonstrate that an attacker who can shift its attack traffic distribution can significantly differ from the original attack traffic and effectively inundate online learning filtering methods. 
Compared to the original attacks, mimic attacks dramatically increase the false positive and false negative error rates.

\vspace{3mm} 
\noindent\textbf{\textit{HULK} attack.}
We compare the online filtering performance on our generated fluctuating attacks and the original attack traffic. As shown in Table~\ref{tab:hulk_compare}, the FPR of the filtering methods significantly increased from less than 10\% to around 60\%. It implies that the filtering systems are systematically classifying normal requests as attacks wrongly, making it worse than a 50\% random guess. 
\begin{table}[htbp]
  \centering
  \caption{HULK shifting traffic. Performance comparison.}
  \resizebox{0.9\linewidth}{!}{%
  \begin{tabular}{c|c|cccccc}
    \hline
    \textbf{Attack} &\textbf{Model} %
    &\textbf{ACC} &\textbf{FPR} & \textbf{Prec.} & \textbf{Rec.} &\textbf{F1} & \textbf{FNR} \\ \hline
    \multirow{4}{*}{\begin{tabular}[c]{@{}c@{}}Original \\ HULK \\ \cite{wjt_asiaccs21} \end{tabular}} &N &0.5903 &0.3837 &0.7016 &0.5739 &0.6313 &--- \\
    &N/D ($\ell=1$)    &0.8086 &0.1030 &0.9198 &0.7524 &0.8277 &--- \\
    &N/D ($\ell=\infty$)   &0.8291 &0.0766 &0.9403 &0.7692 &0.8462 &--- \\
    &Iter.($\ell=1$)  &0.9088	&0.0378	&0.8831	&0.7659	&0.8204 &--- \\ 
    &Iter. ($\ell=\infty$)   &0.9120	&0.0206	&0.9299	&0.7313	&0.8187 &---  \\ \hline
    \multirow{4}{*}{\begin{tabular}[c]{@{}c@{}}Mimic\\ HULK\\ (with $\mathcal{N}$) \end{tabular}} &N  &0.9019 &0.2790 &0.9196    &0.9561  &0.9375 &0.7824 \\ 
    &N/D ($\ell=1$)      &0.7403 &0.6295 &0.8186    &0.8511  &0.8345 &0.8997 \\
    &N/D ($\ell=\infty$) &0.6461 &0.8337 &0.7597    &0.7899  &0.7745 &0.9693 \\
    &Iter. ($\ell=1$)  &0.8435 &0.5684 &0.8780 &0.9252 &0.9010 &0.6802 \\ 
    &Iter. ($\ell=\infty$)   &0.7864 &0.6619 &0.8702 &0.8548 &0.8624 &0.7006 \\ \hline
    \multirow{4}{*}{\begin{tabular}[c]{@{}c@{}}Mimic\\ HULK\\ (without $\mathcal{N}$) \end{tabular}} &N  &0.9045	&0.2734	&0.9212	&0.9578	&0.9392	&0.7834 \\ 
    &N/D ($\ell=1$)      &0.7340	&0.6432	&0.8147	&0.8470	&0.8305	&0.9009 \\ 
    &N/D ($\ell=\infty$) &0.6427	&0.8411	&0.7576	&0.7877	&0.7724	&0.9697 \\ 
    &Iter. ($\ell=1$)  &0.6514	&0.6772	&0.8326	&0.6846	&0.7514	&1.0000 \\ 
    &Iter. ($\ell=\infty$)   &0.7930	&0.4329	&0.8786	&0.8482	&0.8631	&0.2169  \\ \hline
  \end{tabular}
}
\label{tab:hulk_compare}
\end{table}
    
Moreover, using the N model as a baseline, the accuracy of both the online filters N-over-D and Iterative Classifier significantly dropped from 90\% to as low as 65\%. The fluctuating mimic traffic fools the online learning filtering methods. The other metrics, such as Precision, Recall, and F1, are also significantly lower than the N-only model, which serves as our benchmark of a clean model that is not uncorrupted by the attack traffic in its filtering evaluation.

\vspace{3mm} 
\noindent\textbf{\textit{LOIC} attack.}
As for the \textit{LOIC} attack, it can be seen in Table~\ref{tab:loic_compare} that the FPR has increased from values close to zero in the original attack traffic to values ranging from (20--97)\%. The accuracy of the online filters has also dropped from 90\% in the original attack to values less than 50\% on the mimic attacks. 

    \begin{table}[htbp]
  \centering
  \caption{LOIC shifting traffic. Performance comparison.}
  \resizebox{0.9\linewidth}{!}{%
  \begin{tabular}{c|c|cccccc}
    \hline
    \textbf{Attack} &\textbf{Model} %
    &\textbf{ACC} &\textbf{FPR} &\textbf{Prec.} &\textbf{Rec.} &\textbf{F1} &\textbf{FNR} \\ \hline
    \multirow{4}{*}{\begin{tabular}[c]{@{}c@{}}Original\\ LOIC\\ \cite{wjt_asiaccs21} \end{tabular}} &N &0.3257 &0.6997 &0.3889 &0.3454 &0.3659 &--- \\ 
    &N/D ($\ell=1$)   &0.9330 &0.0047 &0.9959	&0.8846	&0.9370 &--- \\ 
    &N/D ($\ell=\infty$)   &0.9330	&0.0047	&0.9959	&0.8846	&0.9370 &--- \\ 
    &Iter. ($\ell=1$)   &0.8805	&0.0329	&0.8808	&0.6494	&0.7476 &--- \\ 
    &Iter. ($\ell=\infty$)   &0.9486	&0.0220 &0.9016	&0.8275	&0.8630 &--- \\ \hline
    \multirow{4}{*}{\begin{tabular}[c]{@{}c@{}}Mimic\\ LOIC\\ (with $\mathcal{N}$) \end{tabular}} &N  &0.9062 &0.2350 &0.9411 &0.9416  &0.9414 &0.7863 \\  
    &N/D ($\ell=1$)      &0.7386 &0.6532 &0.8364 &0.8368  &0.8366 &0.8882 \\  
    &N/D ($\ell=\infty$) &0.6100 &0.9741 &0.7560 &0.7564  &0.7562 &0.9836 \\
    &Iter. ($\ell=1$)   &0.2085 &0.2003 &0.0520 &0.5538 &0.0951 &0.8823 \\ 
    &Iter. ($\ell=\infty$)   &0.3798 &0.4102 &0.3129 &0.7795 &0.4465 &0.6943 \\ \hline
    \multirow{4}{*}{\begin{tabular}[c]{@{}c@{}}Mimic\\ LOIC\\ (without $\mathcal{N}$) \end{tabular}} &N  &0.9028	&0.2434	&0.9390	&0.9395	&0.9393	&0.7869 \\ 
    &N/D ($\ell=1$)      &0.6855	&0.7857	&0.8032	&0.8036	&0.8034	&0.9269 \\ 
    &N/D ($\ell=\infty$) &0.6078	&0.9796	&0.7546	&0.7550	&0.7548	&0.9877 \\ 
    &Iter. ($\ell=1$)  &0.8330	&0.8364	&0.8272	&1.0000	&0.9054	&1.0000 \\ 
    &Iter. ($\ell=\infty$)   &0.3402	&0.2514	&0.8042	&0.2312	&0.3591	&0.9265  \\ \hline
  \end{tabular} 
}
\label{tab:loic_compare}
\end{table}

\vspace{3mm} 
\noindent\textbf{\textit{CAIDA07} attack.}
In the real-world \textit{CAIDA07} attack, the results (see Table~\ref{tab:caida_compare}) corroborate with the findings in both the previous attacks, \textit{LOIC} and \textit{HULK}. The FPR is high (closer to 1.0), while the accuracy is much lower than the N-only model.

    \begin{table}[htbp]
  \centering
  \caption{CAIDA07 shifting traffic. Performance comparison.}
  \resizebox{0.9\linewidth}{!}{%
  \begin{tabular}{c|c|cccccc}
    \hline
    \textbf{Attack} &\textbf{Model} %
    &\textbf{ACC} &\textbf{FPR} &\textbf{Prec.} &\textbf{Rec.} &\textbf{F1} &\textbf{FNR} \\ \hline
    \multirow{4}{*}{\begin{tabular}[c]{@{}c@{}}Original \\ CAIDA07 \\ \cite{wjt_asiaccs21} \end{tabular}} &N &0.6009	&0.2639 &0.8958	&0.5671	&0.6945 &--- \\ 
    &N/D ($\ell=1$)   &0.6944	&0.0251 &0.9900	&0.6242	&0.7657 &--- \\
    &N/D ($\ell=\infty$)   &0.6880	&0.0411	&0.9837	&0.6202	&0.7608 &--- \\ 
    &Iter. ($\ell=1$)   &0.7081	&0.0208	&0.9919	&0.6403	&0.7783 &--- \\
    &Iter. ($\ell=\infty$)   &0.7088	&0.0141	&0.9945	&0.6396	&0.7785 &--- \\ \hline
    \multirow{4}{*}{\begin{tabular}[c]{@{}c@{}}Mimic\\ CAIDA07\\ (with $\mathcal{N}$) \end{tabular}} &N &0.9825 &0.0013 &0.9987  &0.9798  &0.9891 &0.7555 \\  
        &N/D ($\ell=1$)        &0.6884 &0.8025 &0.8149  &0.7995  &0.8071 &0.7622 \\ 
        &N/D ($\ell=\infty$)   &0.6928 &0.7905 &0.8177  &0.8022  &0.8099 &0.9288 \\
        &Iter ($\ell=1$)   &0.8818 &0.1109 &0.8754 &0.9773 &0.9236 &0.9434 \\  
        &Iter ($\ell=\infty$)   &0.9753 &1.0000 &0.9998 &0.9708 &0.9851 &0.9266 \\ \hline
    \multirow{4}{*}{\begin{tabular}[c]{@{}c@{}}Mimic\\ CAIDA07\\ (without $\mathcal{N}$) \end{tabular}} &N  &0.9598	&0.0155	&0.9845	&0.9659	&0.9751	&0.7592 \\ 
    &N/D ($\ell=1$)      &0.9324	&0.0326	&0.9674	&0.9491	&0.9582	&0.7849 \\ 
    &N/D ($\ell=\infty$) &0.9278	&0.0355	&0.9645	&0.9463	&0.9553	&0.7882 \\ 
    &Iter. ($\ell=1$)  &0.8005	&0.9978	&0.8132	&0.9806	&0.8891	&0.0317 \\ 
    &Iter. ($\ell=\infty$)   &0.3140	&0.0654	&0.9255	&0.1727	&0.2911	&0.8881  \\ \hline
  \end{tabular}
}
\label{tab:caida_compare}
\end{table}

While the mimic attacks demonstrate that online learning filters are ill-equipped to handle fluctuating traffic distributions, we discover that the N-only model is able to hold up well against such attacks. As the N-only model trains on normal traffic, the attack traffic does not affect the model, thereby enabling it to differentiate between normal and attack traffic effectively. 

\subsection{Attack Analysis}
\label{sec:atk_analysis}
We further analyze the poisoning attacks, shifting attack traffic generated by \textsc{MimicShift} that eludes online DDoS filters, to better understand the extent of the shifting traffic effects. We study the contrast between the estimated traffic distributions during normal and DDoS periods by analyzing these distributions from an empirical perspective. 

The study was conducted by using the N-over-D approach, which estimates the likelihood that a network request is an attack, given that it is observed in the attack traffic. 
The process takes a two-step process, where the first step trains an N-only model on normal traffic to learn normal traffic characteristics. The next step activates another learning model, D, that takes an online approach to model the attack traffic. The approach then approximates the conditional probability of the traffic by taking $N/D$. Hence, we are able to clearly examine how the attack traffic affects the online training process by studying the difference between models N and online D. 

\begin{figure*}[htbp]
    \centering
    \resizebox{.7\linewidth}{!}{%
        \begin{minipage}[b]{.5\textwidth}
        \centering
        \subfloat[][N only \label{fig:Nonly_distr}]{
            \includegraphics[width=\linewidth]{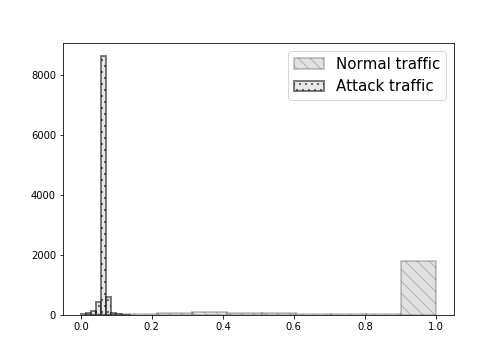}}
        \end{minipage} \qquad \qquad \qquad
        \begin{minipage}[b]{.5\textwidth}
        \centering
            \subfloat[][Online N-over-D \label{fig:NDonline_distr}]{
            \includegraphics[width=\linewidth]{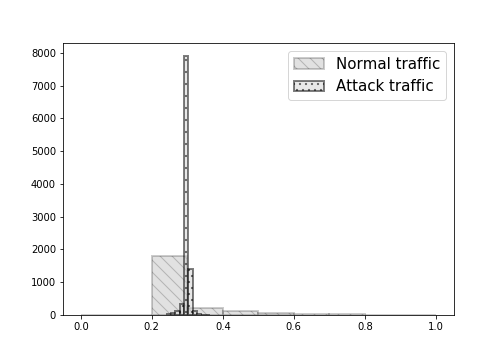}}
        \end{minipage}
    }
    \caption{The estimated distributions of the normal and CAIDA shifting attack traffic determined by models (a) N-only and (b) online N-over-D. The online approach has been poisoned by the fluctuating attack traffic, resulting in the successful attack. The x-axis is the normalized score values (0--1), and the y-axis is the traffic volume.}
    \label{fig:ND_distr}
    \end{figure*}

Using the \textit{CAIDA07} dataset for illustration, we study the normalized conditional probability scores of the attack traffic (see Figure~\ref{fig:ND_distr}). Through these distributions, we observe the differences between an anomaly detection method N-only and the online learning approach N-over-D. 
In Figure~\ref{fig:Nonly_distr}, we plot the scores of normal traffic and attack traffic predicted with the N-only model, showing a clear distinction between the two distributions. It can be seen that the attack traffic is assigned a low score. In contrast, the normal traffic is assigned high scores, indicating that the anomaly detector is able to separate the malicious from the benign. 

However, when we introduce the online model D into the filtering process, the online filtering approach is deeply poisoned by the shifting traffic. We can clearly see in Figure~\ref{fig:NDonline_distr} that the online approach estimates the likelihood of both normal and attack traffic with the same scores, and it is unable to separate the two distributions. 
Hence, \textsc{MimicShift} can effectively elude online DDoS filters by mimicking normal traffic and injecting erratic attack traffic into the online learning process.

\subsection{Countermeasure against Poisoning}
While the results in \cref{sec:eval_results,sec:eval_compare,sec:atk_analysis} demonstrate that poisoning attacks are highly effective against online adaptive filtering systems, we introduce a simple protective measure that greatly diminishes the attack efficacy. It is a straightforward defense technique that can be quickly implemented. 
The idea of the online learning filter enhancement is to vary time interval $\ell$ in the online learning process during the attack period. 

We perform the protective countermeasure by first setting the initial attack interval $\mathcal{A}_0$ to the interval size $\ell=5$ mins. As for subsequent intervals ($\mathcal{A}_1, \mathcal{A}_2, \ldots$), we randomly set $\ell$ to be between 1--3 mins.
By keeping the interval $\ell$ random and unknown to the attacker, online filtering systems are able to improve the estimate of the overall attack traffic distribution while achieving adaptive filtering. 

\vspace{7mm} 
\noindent\textbf{Two-class Iterative Approach.}
The performance of our improved two-class online approach, the Enhanced Iterative Classifier, demonstrates that the introduced countermeasure successfully reduces its FNR and FPR significantly (see Table~\ref{tab:iterperf_enhanced}). 
In addition to the joint training of the Enhanced Iterative Classifier, this countermeasure effectively mitigates against the poisoning attacks.




\begin{table}[htbp]
    \centering
    \caption{Comparison of the iterative classifier performance against both types of \textsc{MimicShift} attacks.}
    \resizebox{0.9\linewidth}{!}{%
        \begin{tabular}{Sc|Sc|Scccccc}
        \hline
        \textbf{Dataset} & \textbf{Model ($\ell=1$)} & \textbf{FNR} & \textbf{FPR} & \textbf{ACC} & \textbf{Prec.} & \textbf{Rec.} & \textbf{F1} \\ \hline 

        \multirow{2}{*}{\begin{tabular}[c]{@{}c@{}}HULK\\ (with $\mathcal{N}$) \end{tabular}} 
        &Iter. Classifier  &0.6802 &0.5684 &0.8435 &0.8780 &0.9252 &0.9010    \\ 
        &Enhanced Iter.  &0.2562 &0.4489 &0.6826 &0.8893 &0.6710 &0.7649	   \\ \cline{2-2}
        \multirow{2}{*}{\begin{tabular}[c]{@{}c@{}}LOIC\\ (with $\mathcal{N}$) \end{tabular}} 
        &Iter. Classifier  &0.8823 &0.2003 &0.2085 &0.0520 &0.5538 &0.0951   \\ 
        &Enhanced Iter. &0.7470 &0.7570 &0.3749 &0.7216 &0.3553 &0.4761   \\ \cline{2-2}
        \multirow{2}{*}{\begin{tabular}[c]{@{}c@{}}CAIDA07\\ (with $\mathcal{N}$) \end{tabular}} 
        &Iter. Classifier &0.9434 &0.1109 &0.8818 &0.8754 &0.9773 &0.9236    \\  
        &Enhanced Iter. &0.1035 &0.0305 &0.9386 &0.9933 &0.9310 &0.9611   \\ \hline

        \multirow{2}{*}{\begin{tabular}[c]{@{}c@{}}HULK\\ (without $\mathcal{N}$) \end{tabular}} 
        &Iter. Classifier  &1.0000	&0.6772	&0.6514	&0.8326	&0.6846	&0.7514 \\ 
        &Enhanced Iter.  &0.7179	&0.4730	&0.4915	&0.8180	&0.4363	&0.5691 \\ \cline{2-2}
        \multirow{2}{*}{\begin{tabular}[c]{@{}c@{}}LOIC\\ (without $\mathcal{N}$) \end{tabular}} 
        &Iter. Classifier  &1.0000	&0.8364	&0.8330	&0.8272	&1.0000	&0.9054 \\  
        &Enhanced Iter. &0.7234	&0.5730	&0.5328	&0.8365	&0.5167	&0.6388 \\ \cline{2-2}
        \multirow{2}{*}{\begin{tabular}[c]{@{}c@{}}CAIDA07\\ (without $\mathcal{N}$) \end{tabular}} 
        &Iter. Classifier &0.0317	&0.9978	&0.8005	&0.8132	&0.9806	&0.8891 \\   
        &Enhanced Iter. &0.9981	&0.9996	&0.3311	&0.9531	&0.1890	&0.3154 \\ \hline         
        
        \end{tabular}
    }
    \label{tab:iterperf_enhanced}
\end{table} 

\vspace{3mm} 
\noindent\textbf{Likelihood Approach.}
As for the principled N-over-D approach, the performance of this countermeasure shows that it achieves similar a performance to the original approach (see Table~\ref{tab:NDperf_enhanced}).




\begin{table}[htbp]
    \centering
    \caption{Comparison of the N-over-D approach performance against both types of \textsc{MimicShift} attacks.}
    \resizebox{0.9\linewidth}{!}{%
        \begin{tabular}{Sc|Sc|Scccccc}
        \hline
        \textbf{Dataset} & \textbf{Model ($\ell=1$)} & \textbf{FNR} & \textbf{FPR} & \textbf{ACC} & \textbf{Prec.} & \textbf{Rec.} & \textbf{F1} \\ \hline 

        \multirow{2}{*}{\begin{tabular}[c]{@{}c@{}}HULK\\ (with $\mathcal{N}$) \end{tabular}} 
        &N-over-D   &0.8997 &0.6295 &0.7403 &0.8186    &0.8511  &0.8345      \\
        &Enhanced N/D  &0.9142 &0.6756 &0.7191 &0.8053 &0.8373 &0.8210	   \\ \cline{2-2}
        \multirow{2}{*}{\begin{tabular}[c]{@{}c@{}}LOIC\\ (with $\mathcal{N}$) \end{tabular}} 
        &N-over-D  &0.8882 &0.6532 &0.7386 &0.8364 &0.8368  &0.8366  \\
        &Enhanced N/D  &0.9312 &0.8085 &0.6764 &0.7975 &0.7979 &0.7977   \\ \cline{2-2}
        \multirow{2}{*}{\begin{tabular}[c]{@{}c@{}}CAIDA07\\ (with $\mathcal{N}$) \end{tabular}} 
        &N-over-D &0.7622 &0.8025 &0.6884 &0.8149  &0.7995  &0.8071       \\ 
        &Enhanced N/D  &0.7622 &0.1851 &0.6884 &0.8149 &0.7995 &0.8071   \\  \hline

        \multirow{2}{*}{\begin{tabular}[c]{@{}c@{}}HULK\\ (without $\mathcal{N}$) \end{tabular}} 
        &N-over-D   &0.9009	&0.6432	&0.7340	&0.8147	&0.8470	&0.8305 \\
        &Enhanced N/D  &0.9164	&0.6936	&0.7108	&0.8001	&0.8319	&0.8157 \\ \cline{2-2}
        \multirow{2}{*}{\begin{tabular}[c]{@{}c@{}}LOIC\\ (without $\mathcal{N}$) \end{tabular}} 
        &N-over-D  &0.9269	&0.7857	&0.6855	&0.8032	&0.8036	&0.8034 \\
        &Enhanced N/D  &0.9385	&0.8244	&0.6700	&0.7935	&0.7939	&0.7937 \\ \cline{2-2}
        \multirow{2}{*}{\begin{tabular}[c]{@{}c@{}}CAIDA07\\ (without $\mathcal{N}$) \end{tabular}} 
        &N-over-D &0.7849	&0.0326	&0.9324	&0.9674	&0.9491	&0.9582 \\
        &Enhanced N/D  &0.7848	&0.0330	&0.9317	&0.9670	&0.9487	&0.9578 \\ \hline 
        
        \end{tabular}
    }
    \label{tab:NDperf_enhanced}
\end{table} 

N-over-D estimates the likelihood that a request is an attack given that it is observed in the attack traffic; the enhanced two-step learning process cannot separate the attack from normal traffic. 
The attack analysis (see Section~\ref{sec:atk_analysis}) helps us better understand why it is challenging for this two-step process to differentiate the attacks.
The attacks generated by \textsc{MimicShift} are designed to closely mimic the normal traffic, resulting in a very high similarity to the normal traffic. 
This approach trains two different models to estimate the conditional probabilities of normal and attack traffic separately. Hence, both attack and normal traffic distributions are close, so it learns to assign them with similar likelihoods (see Figure~\ref{fig:NDonline_distr}). Hence, it becomes tough for this approach to differentiate normal from attack traffic. 

\begin{figure*}[!htb]
    \centering
    \resizebox{0.9\linewidth}{!}{
        \begin{minipage}[b]{.33\textwidth}
        \centering
        \subfloat[][HULK (FNR)]{
            \includegraphics[width=1.0\linewidth]{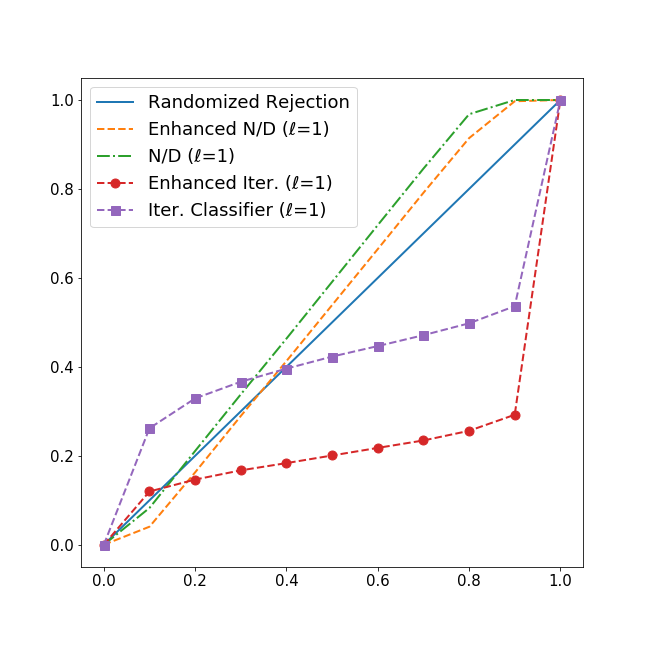}}
        \end{minipage} \qquad
        \begin{minipage}[b]{.33\textwidth}
        \centering
        \subfloat[][LOIC (FNR)]{
            \includegraphics[width=1.0\linewidth]{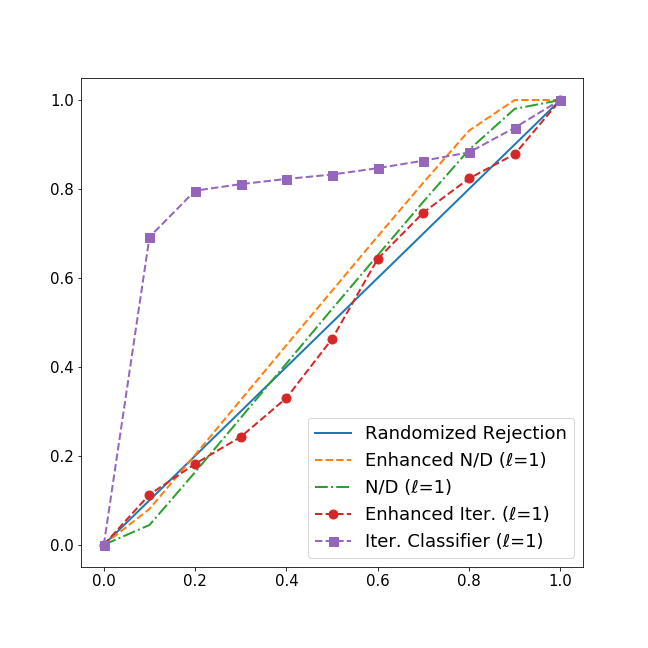}}
        \end{minipage} \qquad 
        \begin{minipage}[b]{.33\textwidth}
        \centering
            \subfloat[][CAIDA07 (FNR)]{
            \includegraphics[width=1.0\linewidth]{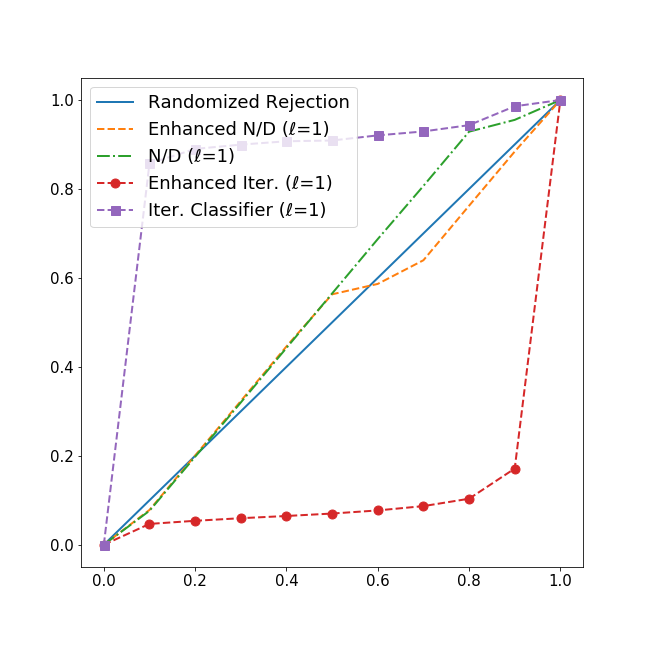}}
        \end{minipage} 
    }
    \caption{The countermeasure greatly improves the performance of the iterative classifier. It can be seen that the false-negative rates of the Enhanced Iterative Classifier are much lower than that of the Iterative Classifier. The x-axis is the acceptance ratio (0--1), and the y-axis is the corresponding false accept rate. 
    }
    \label{fig:fneg_enhanced}
\end{figure*}

In addition, we plot the false-negative graphs of the implemented enhanced countermeasure for all three attacks (see Figure~\ref{fig:fneg_enhanced}). It demonstrates that while the countermeasure slightly improves the performance of the N-over-D approach, it significantly enhances the performance of the Iterative Classifier.

\section{Discussion}
\label{sec:discuss}
In this section, we discuss the current limitations of \textsc{MimicShift} and potential DDoS filtering strategies against non-stationary attack traffic, reviewing the implications.

\subsection{Limitations}
The limitations of \textsc{MimicShift} arise from the design objective, where the generated attack traffic aims to be real with respect to the attacker. The goal of the generative model is for the poisoning attack to work on the online learning filters. It is designed to fool the learning filters and for the defense to fail. Hence, it is a lower requirement, and the fidelity of the generated traffic is not as essential for this requirement. 
In addition, \textsc{MimicShift} requires a reasonable estimate of the interval $\ell$ to shift its traffic accordingly to poison the online learning filters effectively. Finally, there is an inherent speed and performance trade-off in the interval size, as a larger interval size reduces the rate of detection and increases the filtering performance. In contrast, a smaller interval results in faster filtering and reduces filter performance. The attacker is subject to the effects of such changes in the interval sizes.

\subsection{Existing Online Filtering Methods}
Existing approaches to the online filtering of DDoS attacks rely either on simple decision models~\cite{7987186}, rudimentary learning algorithms~\cite{DANESHGADEHCAKMAKCI2020102756,lima2019smart}, or deep learning models~\cite{wjt_asiaccs21} to rank the network requests according to their legitimacy. The filtering methods then set a threshold proportion of the lowest ranking traffic to reject, assuming that the methods assign attack requests a low score accurately and vice-versa for the normal requests. As such, it has been suggested that increasing the threshold to reject a larger proportion of traffic could further mitigate the attacks~\cite{wjt_asiaccs21}. However, our results show that there is a trade-off between increasing the threshold, which increases the FPR and reduces the FNR, rendering the above-mentioned filtering methods ineffective.
For example, a high threshold reduces the number of falsely rejected requests but at the same time falsely classifies more attack traffic as normal, allowing them through to the target server.

\subsection{Potential Filtering Strategies}
\noindent\textbf{Hardening the Accept Threshold.}
The previous deep learning online filtering work suggested that the approaches~\cite{wjt_asiaccs21} are robust against the choice of rejection threshold (0.5--0.7) and time interval $\ell=\{1,2,5,10\}$ for each period of learning the attack traffic. One reason for the robustness is that volumetric DDoS attacks characteristically make up more than half the traffic, reaching up to 80–90\% of the traffic volume, and this large proportion of attack enables a higher threshold to not adversely affect the filtering performance. 

In Tann et al.~\cite{wjt_asiaccs21}, the authors reported that adjusting the rejection threshold from 0.5 to 0.7 into steps of 0.05 has a negligible effect on the false positive rate. However, as we are concerned with attack traffic eluding online filters, we focus on the false-negative rate; our experiments (see Section~\ref{sec:eval_results}) suggest that a change in acceptance threshold could considerably affect the FNR. Thus, a potential filtering strategy would be to set a lower acceptance threshold to increase the chance of stopping attack requests from slipping through the filter. It is a strategy that is worth investigating in future works.
 
\vspace{3mm} 
\noindent\textbf{Random Learning of Attack.} 
One other potential filtering strategy is to set the attack traffic learning model to embed some level of randomness. In particular, the learning strategy of a learning model for attack traffic can choose a time interval $\ell$ from a range of short spans for each period $\mathcal{A}_t$, with some probability. Given that the interval $\ell$ between 1 and 10 mins does not have much of an effect on the filtering performance, the strategy that makes a random selection of $\ell$ for attack traffic learning could possibly mitigate the effects of shifting attack traffic. Further study is needed to analyze the impact of such a countermeasure. 

\vspace{3mm} 
\noindent\textbf{Similarity of Distribution.} 
Another potential filtering strategy for shifting attack distributions is to only accept requests during the attack period with the closest probability score to that of normal traffic. Most filtering methods are able to approximate the normal traffic distribution to a high degree of accuracy. Hence, during an attack, another model can be employed to learn the attack traffic of that specific period and estimate the distribution of the attack requests. With both distributions, the filtering mechanism would only accept requests from the attack traffic that falls in the distribution of normal traffic plus some specified error margin, thereby greatly minimizing the FNR. In other words, the filtering system is designed to be very strict that traffic during a detected attack falls within the scope of normal behavior.

\subsection{Shifting for Improved Filtering}
Optimistically, the controlled synthetic generation of various types of network traffic, most likely unobserved previously, can surprisingly aid in building enhanced DDoS filtering systems robust against poisoning attacks. As this is the first work to suggest poisoning attacks on online deep-learning-based DDoS filtering, the robustness of such filters can be secured by occasionally injecting shifting attack traffic throughout its training process. 
In particular, the generated shifting attack traffic that retains some level of normal traffic properties, especially the ones that are ranked highly to be normal by the online learning systems, can be included and explicitly labeled in the training dataset of the model training during the normal period. We note that by randomly injecting various shifted attack traffic into the normal period training process, the online filters are capable of learning to adapt attack traffic with a wide range of diversity, thereby enhancing the robustness against shifting attack traffic during an attack.

\section{Related Work}
\label{sec:related}
In this section, we examine some of the most relevant works that take statistical and machine learning approaches in DDoS attacks and defenses.

\subsection{Approaches to DDoS Defense}
\noindent\textbf{Statistical.}
Statistical approaches to DDoS mitigation, widely popular when first introduced in the early $2000$s~\cite{6489876}, generally involve methods that measure select statistics of network traffic properties, such as entropy scoring of network packets~\cite{7345297,6601602,8256833,8466805}, applied to IP filtering~\cite{1204223,Peng02detectingdistributed} and IP source traceback~\cite{10.1007/3-540-36159-6_4,5467062}. However, given that DDoS attacks vary from previous attacks, statistical methods are not generally appropriate in an online setting as some traffic statistics cannot be computed during attacks. While Kim et al.~\cite{1354679} proposed a statistical method to support distributed filtering and automated online attack characterization, which assigns a score to each packet that estimates a packet's legitimacy given the attributes, it remains especially challenging for statistical techniques to perform online updates and select appropriate model parameters that minimize false detection rates. 

\vspace{3mm} 
\noindent\textbf{Machine Learning.}
The rapid progress of machine learning has led to the increasingly widespread adoption of deep learning models in DDoS filtering. These learning models can be broadly categorized into two main groups, where one group takes an offline approach~\cite{8066291,10.1007/978-3-030-00018-9_15,doi:10.1002/dac.3497,8416441,Yuan2017DeepDefenseID,8666588}, where the models are trained on all attack data that is observed and available, and the other group adopts an online learning approach~\cite{wjt_asiaccs21,lima2019smart,7987186,DANESHGADEHCAKMAKCI2020102756,Doriguzzi_Corin_2020}, which trains models that only receive data incrementally, as more attack traffic is observed with time.
%
Offline learning methods~\cite{8416441,8066291,10.1007/978-3-030-00018-9_15,doi:10.1002/dac.3497,Yuan2017DeepDefenseID,8666588} utilize various types of deep learning models based on Convolutional Neural Network (CNN) and Recurrent Neural Network (RNN) architectures. It is observed that modeling network traffic through time in a sequential manner, retaining historical information, results in better filtering performance. RNN-based intrusion detection systems~\cite{8066291,10.1007/978-3-030-00018-9_15,doi:10.1002/dac.3497,Yuan2017DeepDefenseID} demonstrate that the sequential learning models demonstrate superior modeling capabilities. They were able to detect sophisticated attacks in various intrusion detection tasks. However, all these methods preprocess the entire training data for model training, and they are not designed for online learning to perform adaptive updates as new data is observed.

\vspace{3mm} 
\noindent\textbf{Online Learning.}
There is only a handful of online learning approaches to DDoS filtering. While an online method~\cite{Doriguzzi_Corin_2020} uses ground-truth labels for training, the other methods~\cite{wjt_asiaccs21,7987186,lima2019smart,DANESHGADEHCAKMAKCI2020102756} do not use data labels for training. However, these approaches either employ basic machine learning algorithms, simple decision tree methods, or even a rudimentary kernel-based learning algorithm with some Mahalanobis distance measure and a chi-square test. Their modeling capabilities limit these online approaches. Tann et al.~\cite{wjt_asiaccs21} presented one of the only approaches that employ powerful online deep learning models, which does not use data labels for training.

\subsection{Attacking DDoS Defenses}
\label{subsec:atkingddos}
There is little literature on adversarial attacks targeted at DDoS detection systems and the corresponding relevant studies. One such analysis~\cite{1498362} was performed to determine an adaptive statistical filter can effectively defend against attackers that dynamically change their behavior. They found that, in fact, the adaptive filter performs much worse against a static attacker than if the attacker is dynamic. While in some recent works~\cite{ibitoye2019analyzing,8466271}, the methods leverage machine learning to generate adversarial samples to fool DDoS detection on deep-learning intrusion detection systems. These attacks generate adversarial samples in a white-box attack setting, where attackers have complete knowledge of the target model. Other methods~\cite{10.1145/3359992.3366642,8995318,8995318,9288358,9344707} present black-box attacks, where the attacker does not know the details of the target model. They perform various manipulative procedures and perturbations on adversarial samples until a sample successfully evades the target DDoS detector. However, there is no study on the effectiveness of poisoning attacks on online deep-learning DDoS filters.

\section{Conclusion}
\label{sec:conclude}
Given the recent success and growing interest in machine learning, we expect increased adoption of online learning models to mitigate and filter prevailing cyber attack issues. We identify a particularly challenging problem, DDoS attacks, a primary concern in present-day network security. 
Hence, we conduct the first systematic study of data poisoning attacks on online deep-learning-based filtering systems. In this study, we show that online learning filters are not robust against ``crafty'' attackers that can shift attack traffic distributions and vary the type of attack requests from time to time. When an online-learning filter wrongly estimates the attack distribution in the next period based on the preceding period, its performance would be much worse than that on a fixed attack distribution. We demonstrate that online adaptive filters are vulnerable to attackers who act in an erratic and unpredictable manner, causing the filter much difficulty to update its adaptation accurately.

In addition, we propose a controllable generative method, \textsc{MimicShift}, for generating attack traffic of specific distributions for poisoning online filtering systems. Such an approach assumes that the generative method only observes the given traffic and has no other knowledge about the filtering systems. The effectiveness of our proposed approach is demonstrated by experimental studies conducted against two recent online deep-learning-based DDoS filtering methods. 
Furthermore, we suggest practical protective countermeasures, which enhance online deep-learning-based defenses against poisoning attacks. Empirical studies verify that our proposed defensive countermeasure can effectively minimize the impact of such attacks, serving as a starting point for the future development of robust defenses.


\bibliographystyle{ACM-Reference-Format}
\bibliography{bibliography}

\appendix
\section*{Appendix}
\subsection*{A. Evaluation Metrics} 
The evaluation metrics are defined as:

\begin{itemize}
\setlength\itemsep{0em}
    \item[] FNR = $FN / (FN + TP)$ 
    \item[] FPR = $FP / (FP + TN)$ 
    \item[] ACC = $(TP + TN) / (TP + TN + FP + FN)$ 
    \item[] Prec. = $TP / (TP + FP)$ 
    \item[] Rec. = $TP / (TP + FN)$ 
    \item[] F1 = $2 [(Prec. \times Rec.) / (Prec. + Rec.)] $
\end{itemize}
where $TP=\text{True Positives}$, $TN=\text{True Negatives}$, $FP=\text{False Positives}$, $FN=\text{False Negatives}$. 
We run all the experiments on NVIDIA Tesla P100 GPUs with 12 GB memory. The models have been implemented in Python v3.7.5 using the Keras v2.2.4 library on top of the Tensorflow v1.14.0 machine learning framework.

\subsection*{B. Feature Analysis} 
We perform feature analysis on the traffic data of the three datasets to analyze the characteristics of the distributions. A large portion of the requests is typically from only a few top classes of the features, concentrating among the top 3--5 classes. 


A summary of the features and the information they provide are listed in Table~\ref{tab:features_decription}. The features include eight dynamic attributes capturing network information that varies from request to request and two static attributes that are common for all requests in a sequence.


\begin{table}[H] 
  \centering
  \caption{Information about the extracted network traffic features. }
    \resizebox{1.0\linewidth}{!}{%
        \begin{tabular}{|Sc|Sc|Sc|}
        \hline
        \textbf{Feature} & \textbf{Information}  &\textbf{Example} \\ \hline
        \begin{tabular}[c]{@{}c@{}}Absolute \\ Time\end{tabular}   &absolute time (in seconds) of the request arrival time  &23557.0 \\ \hline
        \begin{tabular}[c]{@{}c@{}}Request \\ Len\end{tabular}      &size of the request &86 \\ \hline
        IP Flags    &\begin{tabular}[c]{@{}c@{}}IP flags (unsigned integer, 2 bytes) consisting of: \\ Don't fragment, More fragments, Reserved bit, \\ and Security flag.\end{tabular}  &0x00004000  \\ \hline
        TCP Len    &length of the TCP segment  &216.0  \\ \hline 
        TCP Ack    &acknowledgement number (unsigned integer, 4 bytes) &426.0   \\ \hline
        TCP Flags  &flags (unsigned integer, 2 bytes)  &0x00000012  \\ \hline
        TCP Window Size    & \begin{tabular}[c]{@{}c@{}}calculated receive window size allowed \\(unsigned integer, 4 bytes) \end{tabular} &17520.0   \\ \hline
        \begin{tabular}[c]{@{}c@{}}Highest Layer\\ (Protocol)\end{tabular} &last protocol used &MDNS   \\ \hline
        \begin{tabular}[c]{@{}c@{}}Extra \\ Info\end{tabular} &\begin{tabular}[c]{@{}c@{}}extra information on the response of the server \\ after receiving a request\end{tabular} &\begin{tabular}[c]{@{}c@{}}Standard query 0x0000 PTR \\111.2.168.192.in-addr.arpa, \\"QM" question\end{tabular}    \\ \hline
        \begin{tabular}[c]{@{}c@{}}List of \\ Protocols\end{tabular} &the full list of protocols to the highest layer &eth:ethertype:ip:udp:mdns    \\ \hline
        \end{tabular}
    }
\label{tab:features_decription}
\end{table}


Given an interval of traffic, the requests are sorted chronologically, and requests are grouped by source and destination IP addresses and divided into sequences of length 16. The number of classes for each feature of the requests varies across a wide range (see Table~\ref{tab:hulk_featurecount}).

\begin{table}[htbp]
  \centering
  \caption{Number of classes for each extracted network traffic feature in all three datasets. }
    \resizebox{.75\linewidth}{!}{%
        \begin{tabular}{|c|c|c|c|}
        \hline
        \multirow{2}{*}{\textbf{Feature}} & \multicolumn{3}{c|}{\textbf{Num. of classes}} \\ \cline{2-4} 
        & \textit{HULK} & \textit{LOIC} & \textit{CAIDA07}           \\ \hline
        Request Len                 &   3753 &3490 	&1059  \\ \hline
        IP Flags                   &      5 &5 	&2  \\ \hline
        TCP Len                    &   3746 &3463	&1051  \\ \hline
        TCP Ack                    & 827122 &88202 	&7598  \\ \hline
        TCP Flags                  &     15 &12 	&9  \\ \hline
        TCP Window Size            &  11098 &9909 	&1614  \\ \hline
        Protocols                  &    367 &412 	&4  \\ \hline
        Highest Layer(Protocol)    &     42 &42 	&3  \\ \hline
        Info                       &2133226 &498219 &98743  \\ \hline
        TCP Ack                &     84 &72 	&15  \\ \hline
        \end{tabular}
    }
\label{tab:hulk_featurecount}
\end{table}

\subsection*{C. Select Feature Distributions} 
In addition, we select a few different features to plot the distributions and highlight that the traffic is concentrated among only the top few classes of each feature.

\noindent\textbf{IP Flags (\textit{HULK}).}

\begin{table}[H]
  \centering
  \caption{Distributions of the \texttt{IP Flags} feature in the \textit{HULK} dataset. The top three classes make up almost 100\% of the entire dataset.}
\resizebox{.75\linewidth}{!}{%
\begin{tabular}{|c|c|c|c|}
\hline
\textbf{Protocol} &\textbf{Count} &\textbf{Total} &\textbf{Proportion (\%)} \\ \hline
0x00004000       &474183 & 593462  &95.1  \\ \hline
0x00000000       &106706 & 593462 &4.5 \\ \hline
0x00002000       &12570 & 593462  &0.3 \\ \hline
\end{tabular}
}
\label{tab:hulk_ipflags}
\end{table}

\begin{figure}[H]
    \centering
    \includegraphics[width=.7\linewidth]{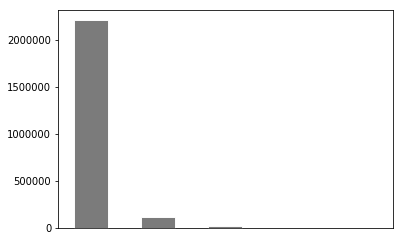}
    \caption{Distribution of the \texttt{IP Flags} feature in the \textit{HULK} dataset.}
    \label{fig:hulk_feature_ipflags}
\end{figure}

\noindent\textbf{TCP Flags (\textit{LOIC}).}

\begin{table}[htbp]
  \centering
  \caption{Distributions of the \texttt{TCP Flags} feature in the \textit{LOIC} dataset. The top five classes make up almost 93.6\% of the entire dataset.}
\resizebox{.75\linewidth}{!}{%
\begin{tabular}{|c|c|c|c|}
\hline
\textbf{TCP Flag} &\textbf{Count} &\textbf{Total} &\textbf{Proportion (\%)} \\ \hline
0x00000010       &421311 & 760674  &55.4  \\ \hline
0x00000018       &134301 & 760674 &17.7 \\ \hline
0                &99479 & 760674  &13.1 \\ \hline
0x00000002       &33459 & 760674  &4.4 \\ \hline
0x00000014       &23183 & 760674  &3.0 \\ \hline
\end{tabular}
}
\label{tab:loic_tcpflags}
\end{table}

\begin{figure}[htbp]
    \centering
    \includegraphics[width=.75\linewidth]{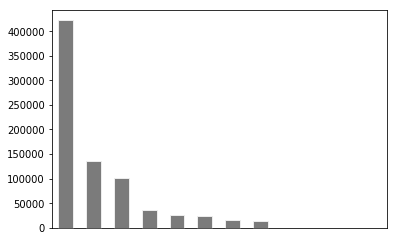}
    \caption{Distribution of the \texttt{TCP Flags} feature in the \textit{LOIC} dataset.}
    \label{fig:loic_feature_tcpflags}
\end{figure}

\noindent\textbf{Protocols (\textit{CAIDA07}).}

\begin{table}[H]
  \centering
  \caption{Distributions of the \texttt{Protocols} feature in the \textit{CAIDA07} dataset. The top three classes make up almost 100\% of the entire dataset.}
\resizebox{.75\linewidth}{!}{%
\begin{tabular}{|c|c|c|c|}
\hline
\textbf{Protocol} &\textbf{Count} &\textbf{Total} &\textbf{Proportion (\%)} \\ \hline
raw:ip:icmp       &474183 & 593462  &79.9  \\ \hline
raw:ip:tcp        &106706 & 593462 &18.0 \\ \hline
raw:ip:icmp:ip    &12570 & 593462  &2.1 \\ \hline
\end{tabular}
}
\label{tab:caida_protocol}
\end{table}

\begin{figure}[H]
    \centering
    \includegraphics[width=.7\linewidth]{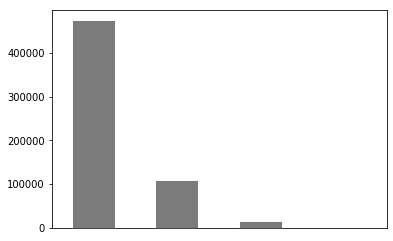}
    \caption{Distribution of the \texttt{Protocols} feature in the \textit{CAIDA07} dataset.}
    \label{fig:caida_feature_protocols}
\end{figure}

\subsection*{D. Implementation Details (\textsc{MimicShift})} 

The \textsc{MimicShift} model consists of a sequence-to-sequence (Seq2Seq) learning model, a generator, and a discriminator.

\noindent\textbf{Mimic Seq2Seq Model.}
In the sequence-to-sequence model, we use an LSTM with 10 cells for all three datasets. The input of this LSTM is a batch of feature sequences length n and dimension d, where the batch size is 128, feature length 16, and dimension is set as the number of classes $\lvert K \rvert = 3$ in each dataset (128 x 16 x d). We select the length as 16 because it should capture most of the typical user behaviors. The LSTM hidden layer with 10 memory units should be more than sufficient to learn this problem. A dense layer with Softmax activation is connected to the LSTM layer, and the generated output is a batch of sequences with size (128 x 16 x d).

\noindent\textbf{Generator.}
In the generator, we use a conditional LSTM with 40 layers
to generate sequences of requests. 
Our generator not only initializes the model with Gaussian noise, but it also takes the features as conditions at each step of the process. 

\begin{figure}[htbp] 
\centering 
\begin{adjustbox}{minipage=\linewidth,scale=0.9}
\begin{empheq}{align*}
    &  &\bm{\bm{z}} \sim & \ \mathcal{N}(\bm{0}, \bm{I}_d) & & \\
    &t=0  &\bm{m}_0 &= g_{\theta'}(\bm{\bm{z}}) & & \\
    &t=1  &f_\theta(\bm{m}_0, \tilde{\bm{s}}_1, \bm{0}) &= (\bm{p}_1, \bm{m}_1),  & \bm{v}_1 &\sim Cat(\sigma(\bm{p}_1)) \\
    &t=2  &f_\theta(\bm{m}_1, \tilde{\bm{s}}_2, \bm{v}_1) &= (\bm{p}_2, \bm{m}_2),  & \bm{v}_2 & \sim Cat(\sigma(\bm{p}_2)) \\
    &\quad\vdots  & &\ \ \vdots & & \qquad \vdots \\
    &t=T  &f_\theta(\bm{m}_{T-1}, \tilde{\bm{s}}_{T}, \bm{v}_{T-1}) &= (\bm{p}_T, \bm{m}_T),  & \bm{v}_T & \sim Cat(\sigma(\bm{p}_T)) 
\end{empheq}
\end{adjustbox}
\end{figure}

Interestingly, we notice that the LSTM generator is more sensitive to the input conditions than hyperparameters for performance. Hence, the set of hyperparameters for the generator is the same for all the datasets. 

\noindent\textbf{Discriminator.}
Our discriminator is an LSTM with 35 layers. The inputs are sequences of requests concatenated with the respective conditions. The discriminator has similar architecture as the mimic model, where they are LSTM models that take sequences as inputs, except that the LSTM layer is connected to a final dense layer. The output is a single value between 0 and 1, which distinguishes real requests from synthetic ones.  

\noindent\textbf{Model Training.}
In the \textsc{MimicShift} training, we use Adam optimizers for all the models. The learning rate of the mimic sequence-to-sequence model training is 0.01, while both the generator and the discriminator use a learning rate of 0.0002.

\subsection*{E. Likelihood Distribution Analysis} 
We include additional attack analysis below to illustrate the similar distributions of the poisoning attacks on both HULK and LOIC datasets. Both figures show that the online likelihood approach assigns both normal and attack traffic with similar scores, and it is unable to differentiate the two distributions, resulting in successful attacks.


\begin{figure}[H]
    \centering
    \includegraphics[width=.75\linewidth]{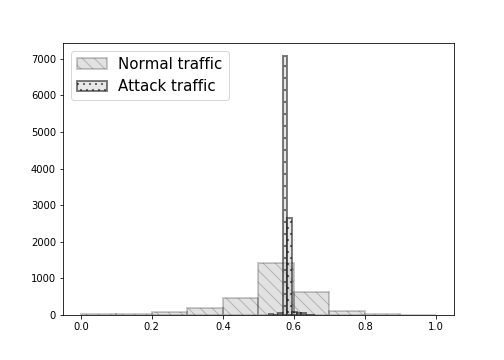}
    \caption{\textsc{MimicShift} attack (HULK). Distributions of the estimated normal and attack traffic.}
    \label{fig:hulk_atkanalysis}
\end{figure}

\begin{figure}[H]
    \centering
    \includegraphics[width=.75\linewidth]{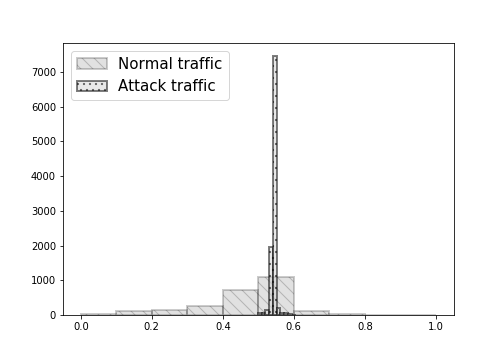}
    \caption{\textsc{MimicShift} attack (LOIC). Distributions of the estimated normal and attack traffic.}
    \label{fig:loic_atkanalysis}
\end{figure}

\subsection*{F. FN and FP Graphs (Attack without $\mathcal{N}$)}

    \begin{figure*}[!htbp]
    \caption{[\textbf{False Negative Rate (Attack without $\mathcal{N}$)}] The x-axis is the acceptance ratio (0--1), and the y-axis is the corresponding false accept rate. If the acceptance ratio is at 0.2, it means that only 20\% of the requests are accepted, and the graph shows the rate of false negatives.}
    \centering
    \resizebox{0.95\linewidth}{!}{
        \begin{minipage}[b]{.33\textwidth}
        \centering
        \subfloat[][HULK (FNR)]{
            \includegraphics[width=1.0\linewidth]{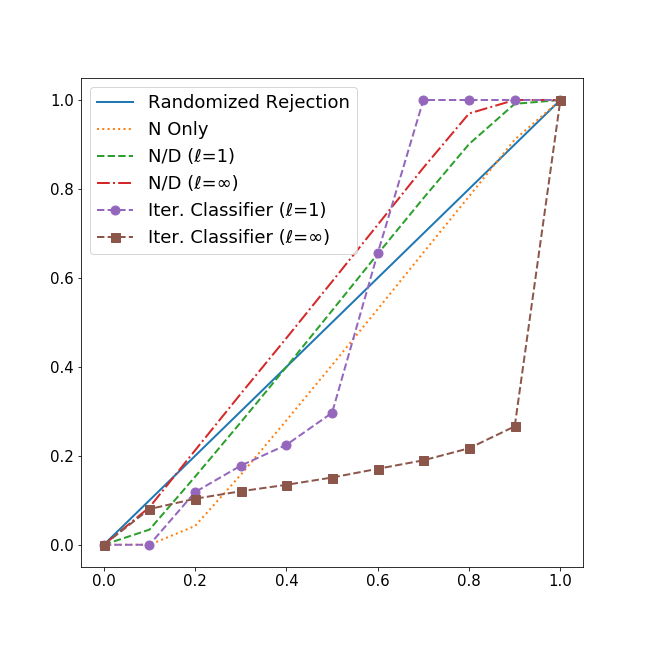}}
        \end{minipage} 
        \begin{minipage}[b]{.33\textwidth}
        \centering
        \subfloat[][LOIC (FNR)]{
            \includegraphics[width=1.0\linewidth]{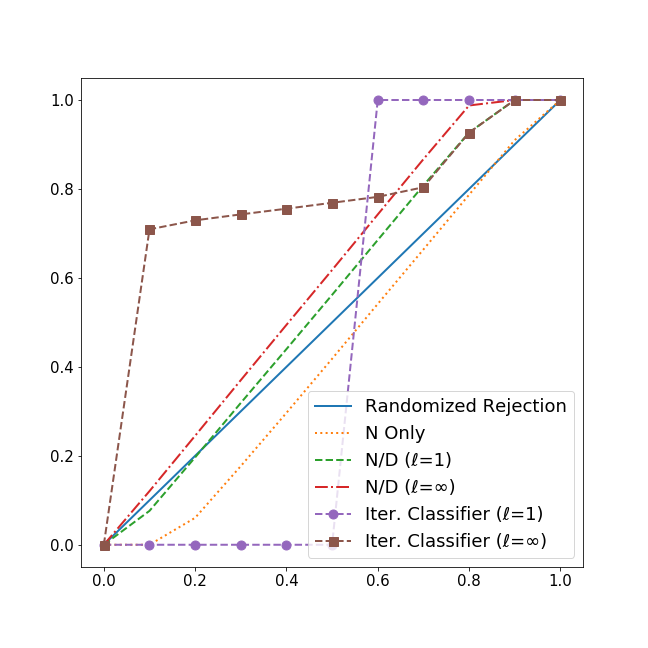}}
        \end{minipage} 
        \begin{minipage}[b]{.33\textwidth}
        \centering
            \subfloat[][CAIDA07 (FNR)]{
            \includegraphics[width=1.0\linewidth]{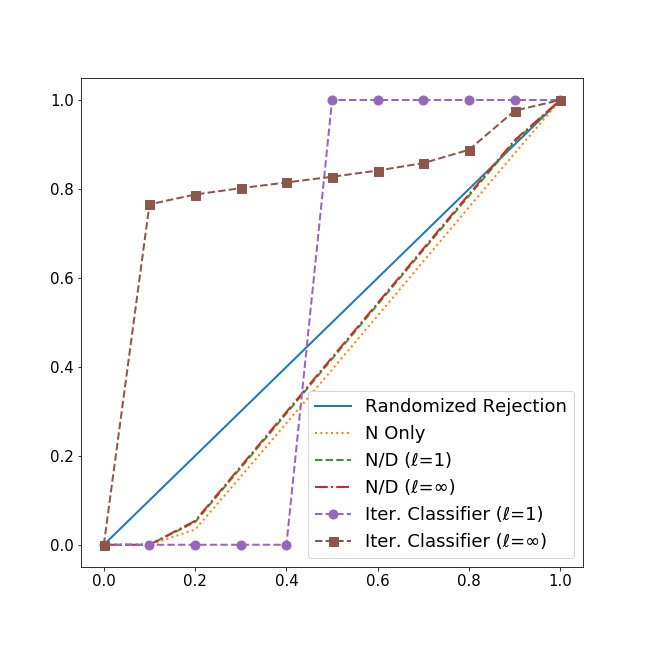}}
        \end{minipage} 
    }
    \label{fig:withoutNfneg_graphs}
\end{figure*} 
    \begin{figure*}[!htbp]
    \centering
    \resizebox{0.95\linewidth}{!}{
        \begin{minipage}[b]{.33\textwidth}
        \centering
        \subfloat[][HULK (FPR)]{
            \includegraphics[width=1.0\linewidth]{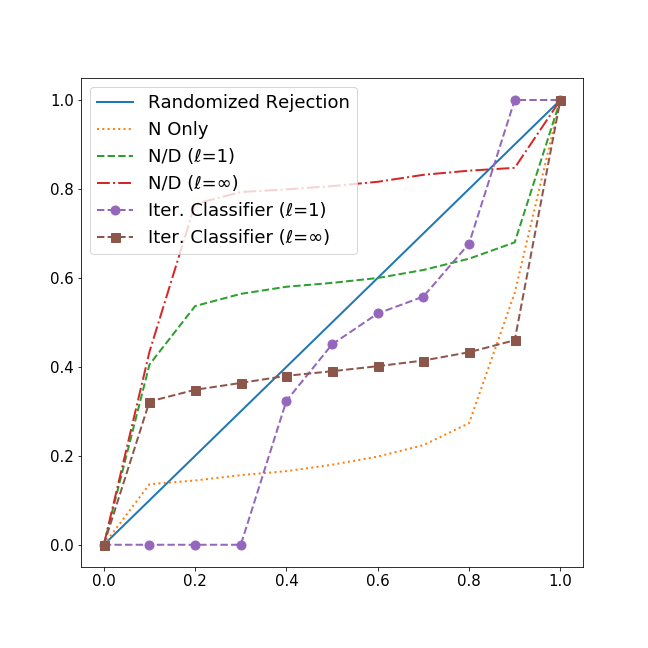}}
        \end{minipage} 
        \begin{minipage}[b]{.33\textwidth}
        \centering
        \subfloat[][LOIC (FPR)]{
            \includegraphics[width=1.0\linewidth]{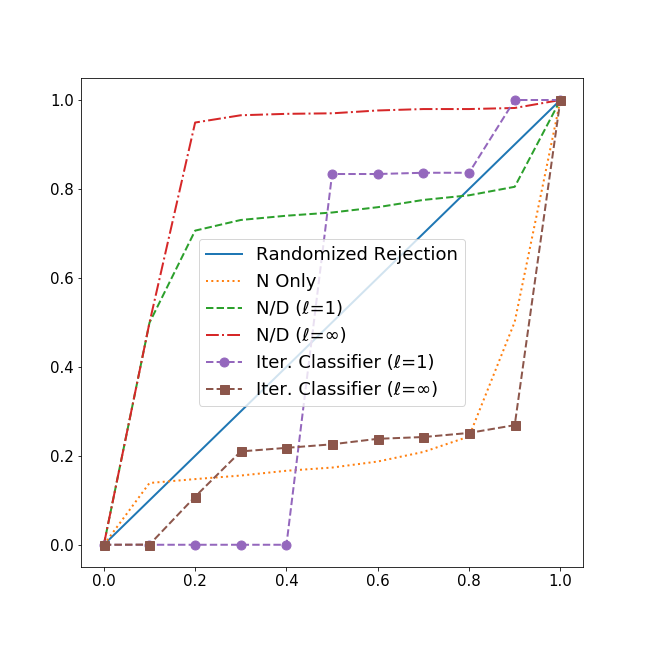}}
        \end{minipage} 
        \begin{minipage}[b]{.33\textwidth}
        \centering
            \subfloat[][CAIDA07 (FPR)]{
            \includegraphics[width=1.0\linewidth]{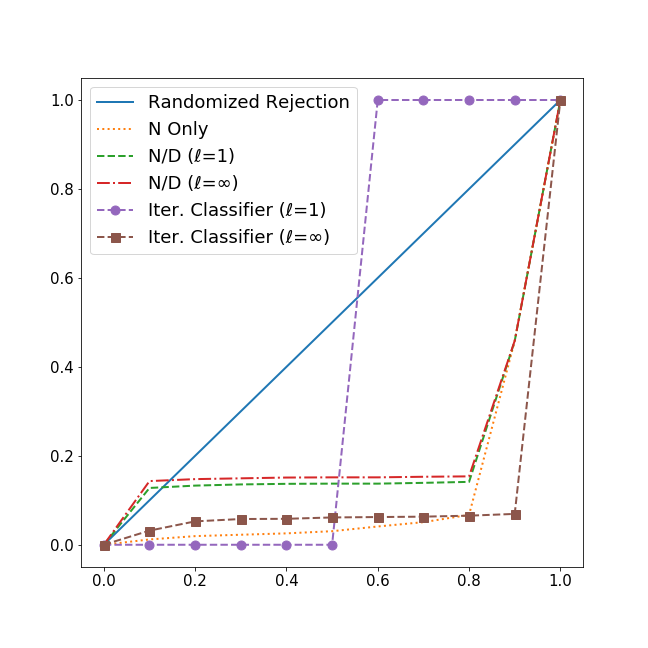}}
        \end{minipage} 
    }
    \caption{[\textbf{False Positive Rate} (Attack without $\mathcal{N}$)] The x-axis is the rejection ratio (0--1), and the y-axis is the corresponding false rejection rate. If the rejection ratio is at 0.8, it means that 80\% of the requests are rejected, and the graph shows the rate of false positives.}
    \label{fig:withoutNfpos_graphs}
\end{figure*} 


\end{document}